\documentclass{aa}  

\usepackage{bm} 
\usepackage{mathrsfs}
\usepackage{graphicx}
\usepackage{txfonts}
\usepackage[colorlinks=true,citecolor=blue,linkcolor=blue,]{hyperref}
\usepackage{multirow}
\usepackage{mathrsfs}

\newcommand{\Eq}[1]{Eq.~(\ref{#1})}

\newcommand{\cP}{c_{\rm P}}
\newcommand{\cV}{c_{\rm V}}

\newcommand{\xxx}{\bm{x}}

\defcitealias{Hidalgo2024}{Paper I}

\begin{document} 


\title{Shaping core dynamos in A-type stars: \\ The role of dipolar fossil fields}
\titlerunning{The role of dipolar fossil fields}

\author{J. P. Hidalgo\inst{1} \and P. J. K\"apyl\"a\inst{2} \and D. R. G Schleicher\inst{1}  \and C. A. Ortiz-Rodríguez\inst{3} \and F. H. Navarrete\inst{4}} 

\institute{Dipartimento di Fisica, Sapienza, Università di Roma, Piazza le Aldo Moro 5, 00185 Roma, Italy \and
Institut f\"ur Sonnenphysik (KIS), Georges-K\"ohler-Allee 401a, 79110 Freiburg, Germany. \and
Hamburger Sternwarte, Universit\"at Hamburg, Gojenbergsweg 112, 21029
Hamburg, Germany. \and 
Institute of Space Sciences (ICE-CSIC), Campus UAB, Carrer de Can Magrans s/n, 08193, Barcelona, Spain.}

   \date{Received XXX; accepted XXX}

 
  \abstract
{Large-scale magnetic fields of Ap/Bp stars are stable over long
  timescales and have typically simple dipolar geometries, leading to
  the idea of a fossil origin. These stars are also expected to have
  convective cores that can host strong dynamo action.}
{We aim to study the interaction between the magnetic fields generated
  by the convective core dynamo of the star, and a dipolar fossil
  field reminiscent of observed magnetic topologies of Ap/Bp stars.}
{We use numerical 3D star-in-a-box simulations of a $2.2M_\odot$ A-type star, where the core encompasses $20\%$ of the stellar radius. As an initial condition, we impose two purely poloidal configurations, both with a surface dipolar strength of 6 kG, and we explore different obliquity angles $\beta$ (the angle between the magnetic and rotational axes), ranging from $0^\circ$ to $90^\circ$.}
{The inclusion of a poloidal field where none of the magnetic field
  lines are closed inside the star, does not affect the core dynamo in
  a significant way. Dipolar configurations where all the field lines
  are closed inside the star can enhance the dynamo, producing a
  superequipartition quasi-stationary solution, where
  the magnetic energy is 5 times stronger than the kinetic energy. The
  enhanced core dynamos have typical magnetic field strengths between
  105 and 172 kG, where the strength has an inverse relation with
  $\beta$. The strong magnetic fields produce an almost rigid rotation
  in the radiative envelope, and change the differential rotation of
  the core from solar-like to anti-solar. The only cases where the
  imposed dipoles are unstable and decay are those with $\beta
  = 90^\circ$. In the rest of cases, the core dynamos are enhanced and
  the surface magnetic field survives keeping
  simple topologies like in the observations.}
{} 


   \maketitle
%

\section{Introduction}

Large-scale magnetic fields have been observed in around $10\%$ of
main-sequence (MS) early-type stars \citep{Moss-2001, Kochukhov-2006,
  Landstreet-2007, Landstreet-2008, Grunhut-2017, Shultz2019}. Unlike
late-type stars, whose magnetic fields have complex geometries and
evolve in relatively short timescales \citep[see
  e.g.][]{2016A&A...595A..12S}, the observed magnetic fields of
early-type stars have simpler geometries, and they are stable
over long timescales, with virtually no variability over several decades 
\citep{2009ARA&A..47..333D,Briquet-2015}. The
observed magnetic fields from late-type stars are very likely to be
due to dynamo processes operating in their convective envelopes
\citep[see e.g.][for the solar dynamo]{Charbonneau-2020}. Numerical
simulations capture many elements of these dynamos \citep[see
  e.g.][and references
  therein]{2023SSRv..219...58K}. However, early-type stars have
radiative envelopes, or very thin convective layers close to the
surface due to peaks in the opacities
\citep[e.g.][]{Richard-2001,Cantiello-2009}. These thin layers are
thought to
produce surface magnetic fields of the order of a few Gauss
\citep{Cantiello-2019}, but this is not sufficient to explain the
observed large-scale magnetic fields of these stars. For example,
chemically peculiar Ap/Bp stars, that
are intermediate mass ($1.5-6~M_\odot$) MS stars, host surface
magnetic fields with mean strengths between 200 G and 30 kG
\citep{2002A&A...392..637S, Auriere-2007}. In most cases the magnetic
field is a simple dipole with its axis misaligned with the rotation
axis \citep{2009ARA&A..47..333D,2015A&A...574A..79K}. Some Ap/Bp stars
exhibit more complex geometries and multipoles are required to
fit the data \citep[e.g.][]{2016A&A...586A..30K,
  2017MNRAS.471..962S}. However, these multipolar fields seem to decay
faster than the commonly observed dipolar fields
\citep{Shultz2019}. As the origin of these large-scale magnetic fields
cannot be explained via convective envelope dynamo, several theories
have been proposed, such as a dynamo operating on the radiative layers
as a result of the interaction of a magnetic instability and
differential rotation. An example is the Tayler-Spruit dynamo
scenario \citep{Spruit-2002, Petitdemange-2023, Petitdemange-2024}.

However, the simple topologies and the stability over long timescales
of the observed magnetic fields seem to support the idea of fossil
fields, that is, magnetic fields originating from an earlier stage of
stellar evolution \citep{2019EAS....82..345A}. In the pre-MS
evolution, Herbig Ae/Be stars \citep{Waters1998,Waters2006} are
believed to be the predecessors of MS Ap/Bp stars, and although their
average magnetic fields are substantially weaker than those from Ap/Bp
stars \citep[see e.g.][]{Hubrig2004, 2019ASPC..518...83K,
  2020AzAJ...15a..68H}, they still share some common characteristics,
such as their incidence rate \citep{2011A&A...536A..45H, Alecian2013,
  2013MNRAS.429.1027A}. This suggests that the magnetic field
is already present in the pre-MS stage, evolving from the molecular
cloud to the MS \citep{Moss2003, Schleicher-2023}. It has also been
proposed that the strong observed magnetic field could be the result
of a merger with another protostar close to the end of the formation
process \citep[e.g.][]{2009MNRAS.400L..71F, 2019Natur.574..211S}. In
any case, the thick radiative envelope of Ap/Bp stars might provide
the ideal conditions for the fossil magnetic field to reach a stable
equilibrium and evolve on a diffusive timescale. \cite{Cowling-1945}
realized that in the radiative core of the Sun, this timescale is of
the order of $10^{10}$ years, therefore, a magnetic field in
equilibrium could survive for the entire MS lifetime of the
star. However, finding stable configurations with analytical methods
has been proven to be a challenging task. The energy method of
\cite{Bernstein-1958} shows that purely poloidal and purely toroidal
fields in highly ideal conditions are unstable under perturbations
\citep{Tayler-1973, Markey-1973, Wright-1973, Markey-1974}. Numerical
MHD simulations by \cite{Braithwaite-2006} produced stable
configurations in the radiative interior of a $2~M_\odot$ A-type star
starting from an initially random field, which relaxed to a stable
roughly axisymmetric torus inside the star, with poloidal and toroidal
components of comparable strength, leading to a roughly dipolar
surface field. Non-axisymmetric configurations have also been found,
starting from turbulent initial conditions
\citep{Braithwaite-2008}. Recently, simulations by \cite{Becerra-2022}
showed that random magnetic field configurations seem to always
evolve to a stable equilibrium in stably stratified environments, 
confirming previous results.

In addition to deep radiative envelopes, early-type stars have
convective cores due to the temperature sensitivity of the CNO
cycle. These cores have vigorous convective
motions and differential rotation that can lead to dynamo
action \citep{Browning-2004}. Indeed, strong core dynamos have been
found in numerical 3D simulations \citep{Brun-2005, Augustson-2016,
  Hidalgo2024}. Recently, an upper
limit of $B_r \approx 500~\mathrm{kG}$ was estimated for the B star HD
43317 by \cite{Lecoanet-2022}, based on its g-mode frequencies
\citep[][]{Buysschaert-2018}. This limit seems to support the idea of
a strong core dynamo inside these stars, due to the fact that this
magnetic field strength would be hard to achieve exclusively with
fossil fields. However, these magnetic fields are most likely unable
to create large-scale structures on the stellar surface
\citep{Schuessler-1978,Parker-1979,MacDonald-2004}, and only a very
small percentage of the magnetic flux can be transported there
\citep{MacGregor-2003, Hidalgo2024}. Interestingly, a core dynamo can
potentially interact with a fossil field \citep[see][in the context of
  the Sun]{Boyer-1984}. \cite{Featherstone-2009} performed numerical
simulations of a $2~M_\odot$ A-type star core dynamo initially hosting
near-equipartition fields surrounded by a fraction of
the radiative envelope. The inclusion of a fossil field with poloidal
and toroidal components led to a superequipartition state where
magnetic energy is roughly ten times stronger than kinetic
energy. This might have some interesting implications, for example, if
this magnetic field is strong enough, then a larger magnetic flux 
can possibly be transported to the surface by magnetic buoyancy
\citep[e.g.][]{MacDonald-2004}. Furthermore, how an enhanced core
dynamo affects the fossil field at the stellar surface and inside the
radiative envelope is unknown.

Simulations studying the stability of magnetic fields assume a
stably stratified environment and neglect the presence of a
core dynamo \citep[e.g.][]{2004Natur.431..819B, Becerra-2022}. In the
current study, we impose a fossil field into a star-in-a-box setup
of a MS A-type star hosting a core dynamo presented in
\cite{Hidalgo2024}; hereafter \citetalias{Hidalgo2024}. This setup
allows us to study how different fossil fields affect the core dynamo
and how the surface magnetic field is modified. Another issue to be
explored, is how
the inclination of the fossil field affects the results, as in most of
the Ap stars the axes of the magnetic dipole and rotation do not
coincide. The models and methods are explained in
Sect.~\ref{Methods}. The initial state of the core dynamo and the
explored fossil fields configurations are described in
Sect.~\ref{imposing}. The results of the simulations are
discussed in Sect.~\ref{results}. Summary and conclusions of the
study are presented in Sect.~\ref{conclusions}.

\begin{figure*}[t!]
    \centering
    \includegraphics[scale=0.37]{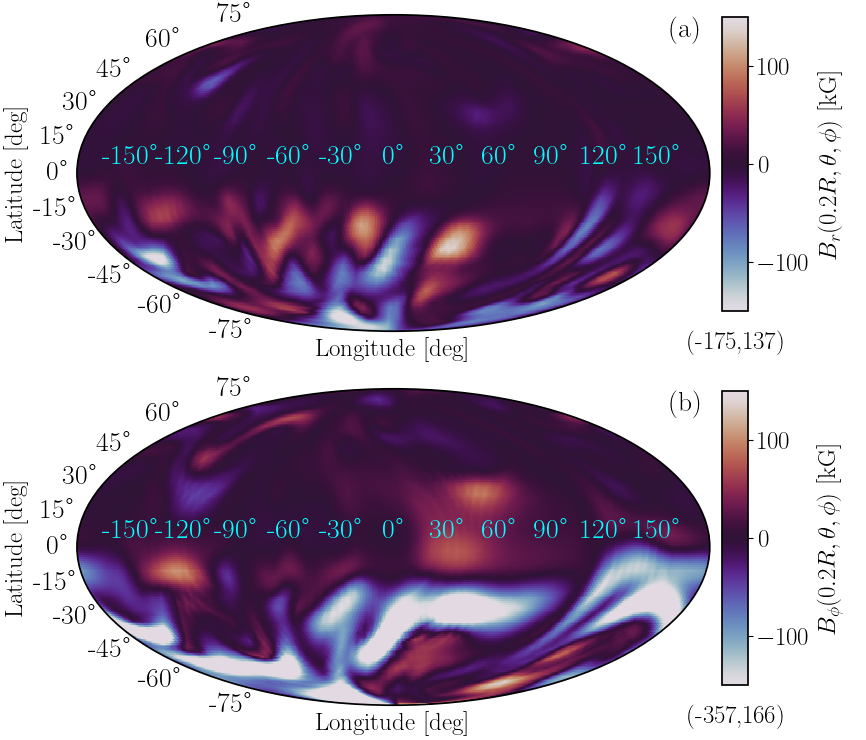}
    \includegraphics[scale=0.383]{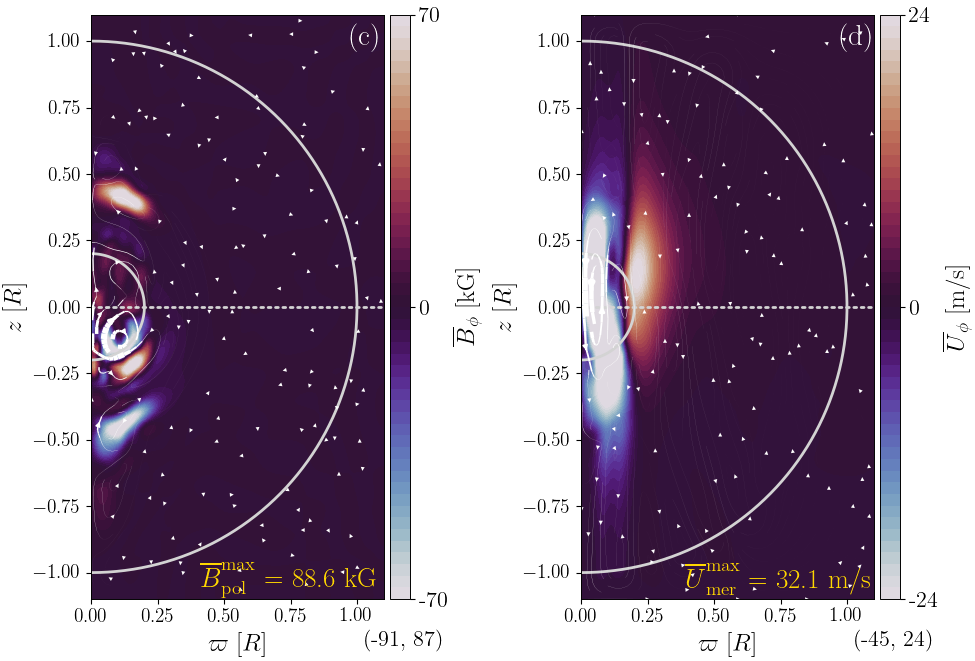}
    \caption{Magnetic fields and flows from a snapshot of MHDr2 at
      $t=85$ yrs. Panels (a), (b): Radial and toroidal magnetic fields at the surface of the convective zone ($r = 0.2R$). Panel (c):  Azimuthally averaged toroidal (colormap) and poloidal (arrows) magnetic fields. Panel (d): Azimuthally averaged toroidal flows (colormap) and meridional circulation (arrows). All the panels are clipped for
      a better display, and the minimum and maximum values are
      indicated below the colorbar.}
    \label{fig:MHDr2-B}
\end{figure*}

\section{Numerical models\label{Methods}}
\subsection{A-type star setup}

The setup used here is the same as that in \citetalias{Hidalgo2024},
which is based on the star-in-a-box
setup from \cite{Kapyla-2021}. The star has a radius $R$ and it is
embedded into a cube of side $l=2.2R$, where the Cartesian coordinates
range from $-l/2$ to $l/2$. The convective core encompasses $20\%$ of
the radial extent of the star, while the rest is stably stratified
\citepalias[see][for more details]{Hidalgo2024}. The fully
compressible set of MHD equations is:
\begin{align}
    \frac{\partial \bm{A}}{\partial t} &= \bm{U} \times \bm{B} - \eta \mu_0 \bm{J}, \label{s3mhd_1} \\
    \frac{D \ln \rho}{D t} &= - \bm{\nabla} \bm{\cdot} \bm{U}, \label{s3mhd_2} \\
    \frac{D \bm{U}}{D t} &=\!-\!\bm{\nabla} \Phi\!-\!\frac{1}{\rho} \left( \bm{\nabla} p\!-\!\bm{\nabla}\!\bm{\cdot}\!2 \nu \rho \bm{\mathsf{S}}\!+\!\bm{J}\!\times\!\bm{B} \right)\!-\!2 \bm{\Omega}\times \bm{U}\!+\!\bm{f}_d, \label{s3mhd_3} \\
    T \frac{Ds}{Dt} &= - \frac{1}{\rho} \left[ \bm{\nabla} \bm{\cdot} (\bm{F}_\text{rad} + \bm{F}_\text{SGS})  + \mathcal{H} - \mathcal{C} + \mu_0 \eta \bm{J}^2 \right] + 2 \nu \bm{\mathsf{S}}^2, \label{s3mhd_4}
\end{align}
where $\bm{A}$ is the magnetic vector potential, $\bm{U}$ is the flow
velocity, $\bm{B} = \bm{\nabla} \times \bm{A}$ is the magnetic field,
$\eta$ is the magnetic diffusivity, $\mu_0$ is the magnetic
permeability of vacuum, $\bm{J} = \bm{\nabla} \times \bm{B}/\mu_0$ is
the current density, $D/Dt = \partial/\partial t
+ \bm{U} \bm{\cdot}\bm{\nabla}$ is the advective
derivative, $\rho$ is the mass density, $p$ is the pressure, $\Phi$ is
the fixed gravitational potential obtained from a 1D model, $\nu$ is the kinematic viscosity, $\bm{\mathsf{S}}$ is the traceless
rate-of-strain tensor
\begin{equation}
  \mathsf{S}_{ij} = \frac{1}{2}(\partial_j U_i + \partial_i U_j) - \frac{1}{3}\delta_{ij} \bm{\nabla} \bm{\cdot} \bm{U}, \label{S-tensor}
\end{equation}
where $\delta_{ij}$ is the Kronecker
delta. The rotation vector is $\bm{\Omega}=\Omega_0 \bm{\hat{z}}$ and $\bm{f}_\mathrm{d}$ damps the flows outside the star, with
\begin{equation}
    \bm{f}_d = - \frac{\bm{U}}{\tau_{\text{damp}}} f_e(r), \label{damping}
\end{equation}
where $\tau_{\text{damp}} = 0.2\tau_\mathrm{ff} \approx 
1.5~\mathrm{days}$ is the damping timescale, $\tau_\mathrm{ff} =
\sqrt{R^3/GM}$ is the freefall time, and
\begin{equation}
    f_e(r) = \frac{1}{2} \left(1 + \tanh\frac{r - r_{\text{damp}}}{w_{\text{damp}}} \right), \label{f_e-damping}
\end{equation}
where $r = \sqrt{x^2+y^2+z^2}$, $r_{\text{damp}}= 1.03R$ is the radius
where the damping starts, and where $w_{\text{damp}}=0.02R$ is its
width. $T$ is the temperature, $s$ is
the
specific entropy, and $\bm{F}_\mathrm{rad}$ is the radiative flux
\begin{align}
    \bm{F}_\mathrm{rad} = - K \bm{\nabla} T, \label{radiative-flux}
\end{align}
where $K$ is the heat conductivity, and $\bm{F}_\mathrm{SGS}$ is the
sub-grid-scale (SGS) entropy flux, which damps entropy fluctuations near 
the grid scale \citep[see][]{Kapyla-2021,2023SSRv..219...58K}
\begin{equation}
    \bm{F}_\mathrm{SGS} = -\chi_\mathrm{SGS} \rho \bm{\nabla}s', \label{entropy-flux}
\end{equation}
where $\chi_\mathrm{SGS}$ is the SGS diffusion coefficient, $s' = s - \langle s \rangle_t$ is the fluctuating entropy, and $\langle s \rangle_t(\xxx)$ is a temporal mean of the specific entropy. 
Because the SGS diffusion applies to the fluctuations of the
  specific entropy, its contribution to net energy transport is
  negligible.
The heating function $\mathcal{H}$ follows
\begin{equation}
    \mathcal{H}(r) = \frac{L_{\text{sim}}}{(2\pi w_L^2)^{3/2}} \exp \left( - \frac{r^2}{2 w_L^2} \right),
\end{equation}
which is a normalized Gaussian that parameterize the
nuclear energy production in the core of the star, where
$L_{\text{sim}}$ is the luminosity in the simulation, and $w_L = 0.1R$
is the width of the Gaussian. The cooling function
$\mathcal{C}(\bm{x})$ is given by
\begin{equation}
    \mathcal{C}(\bm{x}) = \rho c_\text{P} \frac{T(\bm{x}) - T_{\text{surf}}}{\tau_{\text{cool}}}f_e(r),
\end{equation}
which models radiative losses above the stellar surface,
where $\cP$ is the heat capacity at constant pressure,
$T_{\text{surf}}=T(R)$ is the temperature at the stellar surface,
$\tau_{\text{cool}} = \tau_{\text{damp}}$ is a cooling timescale, and
$f_e(r)$ is given by \Eq{f_e-damping}, with $r_{\text{cool}} =
r_{\text{damp}}$ and $w_{\text{cool}} =
w_{\text{damp}}$. The ideal gas equation of state $p = (\cP-\cV)\rho
T$ is assumed, where $\cV$ is the heat capacity at constant
volume. The model has radial profiles for the diffusivities $\nu$ and
$\eta$, where radiative zones have values that are $10^2$ times
smaller than in convective zones \citep{Kapyla-2202}. Furthermore,
sixth-order hyperdiffusivity terms are added in the
dynamical equations \cite[see e.g.][]{Brandenburg-2002-V2,
  Johansen-2005, Lyra-2017}.

The simulations were run with the {\sc Pencil
  Code}\footnote{\url{https://pencil-code.org/}}
\citep{Pencil-code-2021}, which is a highly modular high-order
finite-difference code for solving ordinary and partial differential
equations, on a grid of $200^3$ equally distributed grid points.

\begin{figure*}[t!]
    \centering
    \includegraphics[scale=0.36]{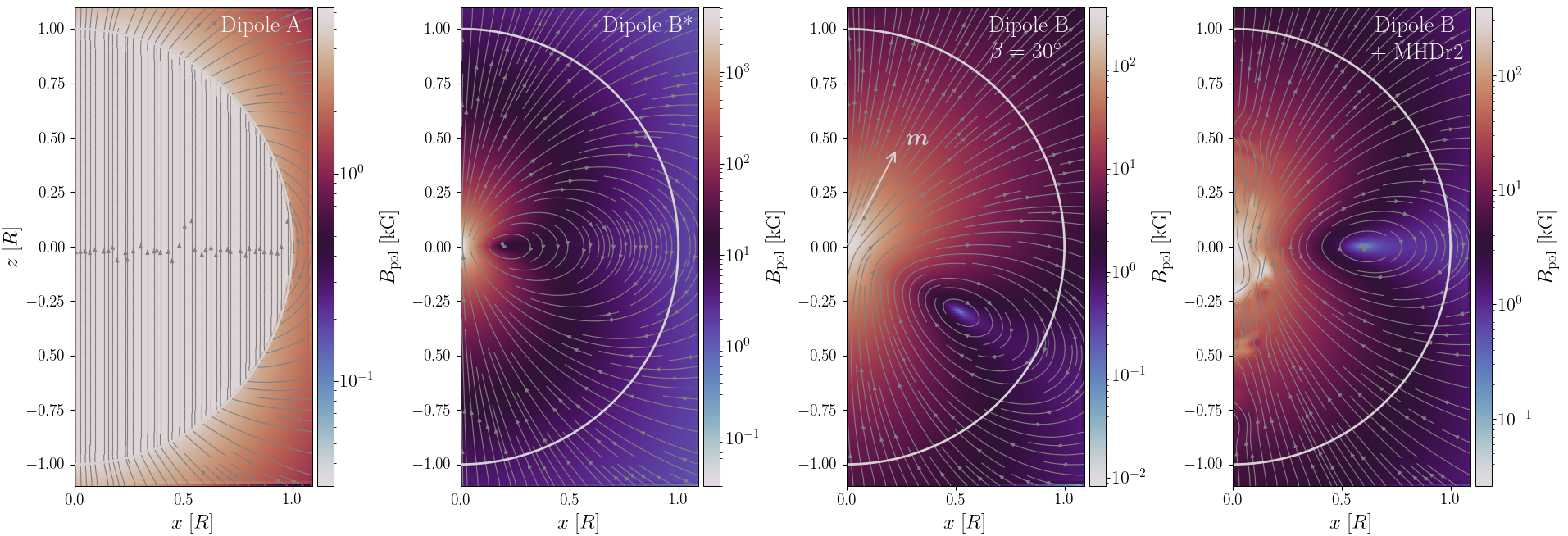}
    \caption{Imposed initial (fossil) magnetic fields in our simulations. The arrows represent the poloidal magnetic field lines, and the colormap the intensity of this component. Dipole A and Dipole B* are aligned to the rotational axis ($\beta=0^\circ$), and as an example of a misaligned dipole an inclination of $\beta = 30^\circ$ was added in Dipole B. The last panel corresponds to the initial snapshot of run DipB (Dipole B + MHDr2).}
    \label{fig:Binit}
\end{figure*}

\subsection{Units and relation to reality}

The units of length, time, density, entropy and magnetic field are
given by
\begin{eqnarray}
    [x]=R, \  [t] = \tau_\mathrm{ff}, \ [\rho] = \rho_0, \ [s] = \cP,  \ [B]= \sqrt{\mu_0 \rho_0}[x]/[t].
\end{eqnarray}
In this study, the quantities are typically expressed in physical
units. The conversion factors are the same as those given by Eqs. (17)
and (18) in \citetalias{Hidalgo2024}. The stellar parameters
where obtained from a 1D MESA model with $2.2~M_\odot$
\citep[][]{Paxton-2019}, and are the same as those from
\citetalias{Hidalgo2024}, i.e., the radius, mass density and
temperature of the stellar center, and stellar luminosity are
$R_\star=2.1~R_\odot$, $\rho_0 = 5.5\cdot 10^4~\mathrm{kg\,m^{-3}}$,
$T_0 = 2.3\cdot 10^{7}~\mathrm{K}$, $L_\star = 23.5~L_\odot$,
respectively. Furthermore, as we are using the fully compressible
formulation of the MHD equations, the enhanced luminosity approach is
also used in these simulations \citep[see][for a detailed
  justification]{Dobler-2006, Kapyla-2020,Kapyla-2021}.
To have a consistent rotational influence on the flow, the rotation
rate is enhanced following the recipe in Appendix~A of
\cite{Kapyla-2020}.

\subsection{Diagnostics quantities}

The rotational influence on the
flow is measured by the global Coriolis number,
\begin{equation}
    \mathrm{Co} = \frac{2 \Omega_0}{u_\mathrm{rms} k_R},
\end{equation}
where $u_\mathrm{rms}$ is the root-mean-square (rms) velocity averaged
over the convection zone and $k_R = 2\pi/\Delta r$ is an estimate of
the largest convective eddies in the system, where $\Delta r = 0.2R$
is the depth of the convective zone. The fluid and magnetic Reynolds
numbers, and the SGS P\'eclet number are defined as:
\begin{eqnarray}
    \mathrm{Re} = \frac{u_\mathrm{rms}}{\nu k_R},\ \ \mathrm{Re_M} = \frac{u_\mathrm{rms}}{\eta k_R},\ \ \mathrm{Pe} = \frac{u_\mathrm{rms}}{\chi_\mathrm{SGS} k_R}.
\end{eqnarray}

\section{Imposing a magnetic field \label{imposing}}
\subsection{Initial state of the simulation}

To analyze how an imposed fossil field affects the core dynamo, the
fossil field configurations are added to snapshots of an already
saturated core dynamo simulation. We chose run MHDr2 from
\citetalias{Hidalgo2024}, which hosts a dynamo with
the largest rotation period ($P_\mathrm{rot}=15$ days) and therefore,
the smallest Coriolis number ($\mathrm{Co} = 10.1$) in that study. The
core of this simulation ($r<0.2R$) has a root-mean-square (rms)
magnetic field of $B_\mathrm{rms} = 60$ kG, solar-like differential
rotation, and a hemispherical dynamo with most of the activity
located on its southern hemisphere. The snapshot where the fossil
field was imposed corresponds to a simulated time of
$t=4135\tau_\mathrm{ff}$, or 85 years in physical units. In
Fig.~\ref{fig:MHDr2-B}(a,b) the radial and
toroidal magnetic fields at the surface of the convective zone
($r=0.2R$) from this snapshot are shown. Both magnetic field
components are highly concentrated on the
southern hemisphere. Furthermore, the amplitude of the toroidal
component is larger than that of the radial component. Panel (c)
displays the azimuthally averaged magnetic field in the entire
box. The hemispheric nature of the core dynamo is still clearly
visible from both components. However, strong
magnetic fields are also present at the bottom radiative
envelope. This is most likely due to flows penetrating into the
radiative layer near the rotation axis. Finally, on panel (d) the
azimuthally
averaged flows are shown.
The flows penetrating deeply into the radiative envelope are most
likely due to lower Brunt-Väisälä frequency \citep[for example, see Fig.~5 of][]{Lydia-2024}, and therefore, lower
Richardson numbers related to rotation than in real stars
\citep[e.g.][]{2024A&A...683A.221K} achieved in our simulations. 
These flows also transport some of the
magnetic flux from the core dynamo to the stellar surface, leading to
an rms surface magnetic field of $\sim 0.1$ kG at the poles and
as low as $10^{-5}$ kG elsewhere.
MHDr2 has subequipartition magnetic fields, where the ratio of the
magnetic to kinetic energies is $E_\mathrm{mag}/E_\mathrm{kin}=0.196$ in the convective core ($r < \Delta r$), and 0.322 in the entire star ($r< R$).

\subsection{Dipolar fossil field \label{dipolar-fossil}}

The majority of Ap stars exhibit large-scale poloidal magnetic
fields. For example, in HD 75049, 90\% of magnetic energy is
concentrated in the dipolar $\ell=1$ spherical harmonic mode
with a polar strength of $B_p=36$ kG and an obliquity
of $36^\circ$ \citep{2015A&A...574A..79K}. Similarly, in 
HD 24712, $96\%$ of the magnetic energy is concentrated in
$\ell=1$ \citep{Rusomarov2015}. Therefore, we choose to impose a
purely poloidal ($B_\phi = 0$) fossil field, whose vector potential is
\begin{equation}
    \bm{A}(\bm{r}) = \frac{\mu_0}{4\pi} \frac{\bm{m} \times \bm{r}}{r^3}, \label{A-dipole}
\end{equation}
where $\bm{m}$ is the magnetic dipole moment. As discussed, the
observed magnetic fields of Ap/Bp stars are typically misaligned with
their rotation axis. Therefore, we define $\bm{m}$ as
\begin{equation}
    \bm{m}(\beta) = m_0(\sin \beta \bm{\hat{x}} + \cos \beta \bm{\hat{z}}), \label{m-tilted}
\end{equation}
where $\beta$ is the obliquity angle of the magnetic dipole with
respect to $\bm{\Omega}$, and $m_0$ is the initial dipolar
amplitude. Thus, Eq.~(\ref{A-dipole}) can be expressed as
\begin{equation}
  \bm{A}(\bm{r},\beta)\!=\!\frac{\mu_0 m_0}{4\pi}\left( - \frac{ y \cos \beta}{r^3} \bm{\hat{x}}\!+\!\frac{x\cos \beta\!-\!z \sin \beta}{r^3} \bm{\hat{y}}\!+\!\frac{y \sin \beta}{r^3} \bm{\hat{z}} \right). \label{A-dipole-inclined}
\end{equation}
The sample of Ap stars presented by \cite{2000A&A...359..213L} shows
that slow rotators ($P_\mathrm{rot} > 25$ days) have small obliquity
angles, with an excess around $20^\circ$, while fast rotators tend to
have larger values of $\beta$. More recently, \cite{Shultz2019}
reported that the distribution of $\beta$ seems to be statistically
consistent with a random distribution, which fits with the data of
\cite{Auriere-2007}, and \citet[][see their
  Fig.~11]{2019MNRAS.483.3127S}. In this study, we consider $0^\circ
\leq \beta \leq 90^\circ$. More specifically, $\beta = 0^\circ~ (0)$,
$30^\circ ~(\pi/6)$, $60^\circ ~(\pi/3)$ and $90^\circ ~(\pi/2)$, to
explore the entire range of inclinations. We further test $\beta =
80^\circ$ ($4\pi/9$) and $85^\circ$ ($17\pi/36$) in one specific case,
to study the transition to a horizontal dipole in more detail.

We define two dipolar initial conditions, the first one
(Dipole A) corresponding to the magnetic field of a magnetized sphere,
that is, a uniform magnetic field inside $(r < R)$, matched by a dipole
field in the exterior $(r > R)$, while the second configuration
corresponds to a dipole field in the entire
domain. Eq.~(\ref{A-dipole-inclined}) has a singularity for $r\to 0$,
and therefore we added a constant $\epsilon$ to $r$, such that
$r' \equiv r + \epsilon$ is used in the calculation of the dipole
fields. Two values of
$\epsilon$ were chosen, where $\epsilon_1 < \epsilon_2$. $\epsilon_1$
($\epsilon_2$) produces a $B_\mathrm{rms}^\mathrm{pol}$ of roughly 500
kG (100 kG) in the core, $r < 0.2R$, and 48 kG (14 kG) in the entire
star. The runs with $\epsilon_1$ are labeled as Dipole B* and those
with $\epsilon_2$ as Dipole B. The initial value of $m_0$ was chosen
such that all the mentioned magnetic configurations have a dipolar
amplitude of roughly 6 kG at the stellar surface, which is typical
from Ap/Bp stars \citep{Auriere-2007, Shultz2019}. All of the
initial conditions are displayed in the first three panels of
Figure~\ref{fig:Binit}. An obliquity angle of $30^\circ$ was added in
Dipole B (third panel) as an example of a tilted dipole initial
condition.



\begin{table*}[h!]
\centering
\caption{Summary of the simulations.}
\begin{tabular}{lcccccccccccc} 
\hline\hline\noalign{\smallskip}
Run  & $\beta$~[$^\circ$] & $u_\mathrm{rms}$~[m/s] & $(B_\mathrm{rms},B_\mathrm{rms}^\mathrm{star}) $~[kG] & $(B_\mathrm{rms}^\mathrm{pol},B_\mathrm{rms}^\mathrm{tor}) $~[kG]  & Co &  Pe & Re & $\mathrm{Re}_\mathrm{M}$ & $P_\mathrm{cyc}$~[years]\\
\hline\noalign{\smallskip} 
MHDr2 & -  &     51  &     $(60,21)$  &  $(    19 ,    22 )$ &   10.1  &      7  &     35  &     24 & $1.88  \pm  0.09$ \\
\hline
DipA & $0^\circ$  &     50  &     $(60,33)$  &  $(    23 ,    19 )$ &    8.7  &      8  &     40  &     28 & $1.66  \pm  0.07$ \\
DipAt & $30^\circ$  &     50  &     $(58,31)$  &  $(    22 ,    19 )$ &    8.7  &      8  &     40  &     28 & $1.79  \pm  0.03$ \\
DipAt2 & $60^\circ$  &     51  &     $(59,28)$  &  $(    20 ,    21 )$ &    8.7  &      8  &     40  &     28 & $1.81  \pm  0.05$ \\
DipAh & $90^\circ$  &     53  &     $(58,23)$  &  $(    18 ,    21 )$ &    8.3  &      8  &     42  &     29 & $1.88  \pm  0.03$ \\
\hline
DipB* & $0^\circ$  &     28  &    $(172,53)$  &  $(   102 ,    54 )$ &   15.6  &      4  &     22  &     15 & no cycles \\
DipBt* & $30^\circ$  &     28  &    $(172,53)$  &  $(    96 ,    65 )$ &   15.8  &      4  &     22  &     15 & no cycles \\
DipBt2* & $60^\circ$  &     31  &    $(141,44)$  &  $(    77 ,    45 )$ &   14.2  &      5  &     24  &     17 & no cycles \\
DipBt3* & $80^\circ$  &     35  &    $(116,37)$  &  $(    54 ,    32 )$ &   12.4  &      5  &     28  &     19 & no cycles \\
DipBt4* & $85^\circ$  &     36  &    $(105,34)$  &  $(    44 ,    27 )$ &   11.9  &      6  &     29  &     20 &  no cycles \\
DipBh* & $90^\circ$  &     52  &     $(53,26)$  &  $(    16 ,    16 )$ &    8.4  &      8  &     42  &     29 & $1.71  \pm  0.26$ \\
\hline
DipB & $0^\circ$  &     29  &    $(157,47)$  &  $(    90 ,    52 )$ &   15.2  &      4  &     23  &     16 & no cycles \\
DipBt & $30^\circ$  &     30  &    $(147,46)$  &  $(    83 ,    49 )$ &   14.5  &      5  &     24  &     17 & no cycles \\
DipBt2 & $60^\circ$  &     33  &    $(129,40)$  &  $(    64 ,    35 )$ &   13.3  &      5  &     26  &     18 & no cycles \\
DipBh & $90^\circ$  &     53  &     $(55,26)$  &  $(    18 ,    20 )$ &    8.5  &      8  &     41  &     29 & $1.66  \pm  0.04$ \\
\hline
\end{tabular} 
\tablefoot{From left to right: obliquity
  angle of the added dipole field, the rms velocity $
  u_\mathrm{rms}$, the volume-averaged rms magnetic field
  $B_\mathrm{rms}$, the poloidal and toroidal components of
  $B_\mathrm{rms}$, the Coriolis number, the SGS P\'eclet number, the
  fluid and magnetic Reynolds numbers, and the magnetic cycle period.
  Quantities are averaged over the convective core $r < 0.2R$ unless
  “star” is indicated, in which case the average is over $r < R$.}
\label{table-results}
\end{table*}

\section{Results \label{results}}

The simulations are divided into three sets, the names of which
reflect the chosen initial configuration. For example, runs with
Dipole B are labeled
as DipB. The simulations, as well as the diagnostic quantities, are
listed in Table~\ref{table-results}.

\subsection{Core dynamos}

\begin{figure}[t!]
    \centering
    \includegraphics[width=\hsize]{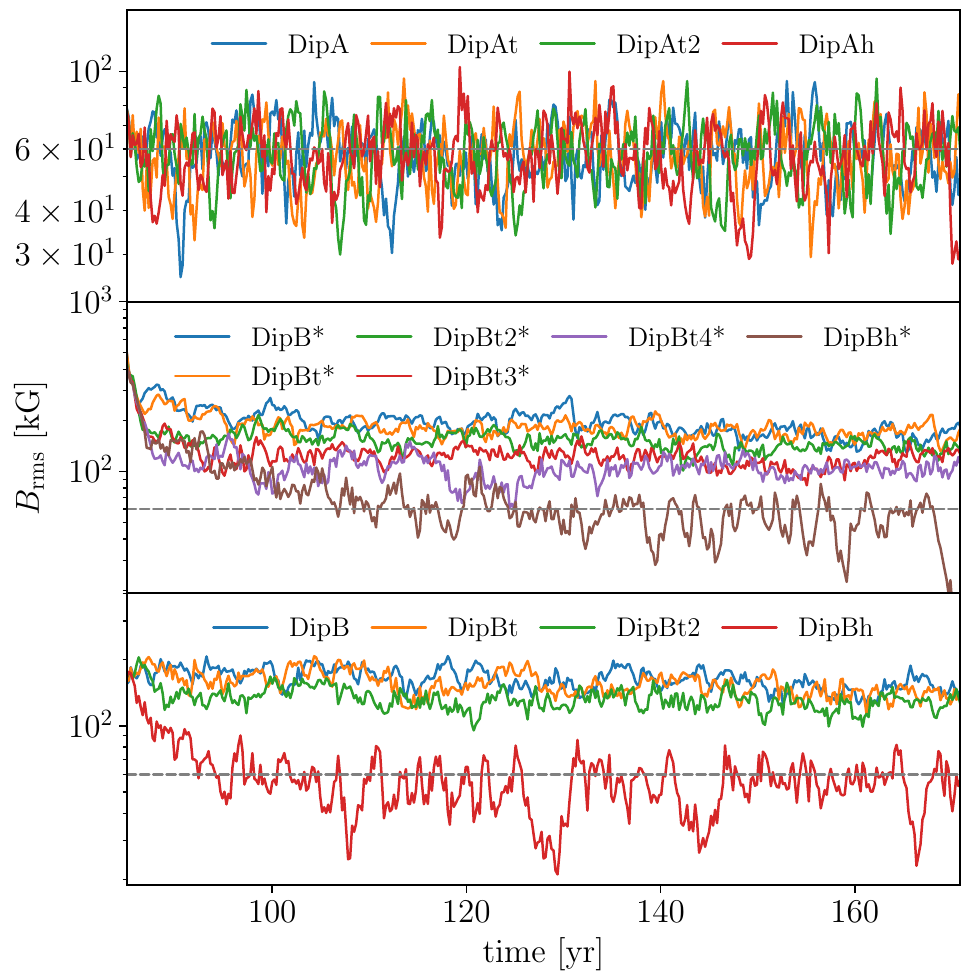}
    \caption{Temporal evolution of the rms magnetic
       field in the convective core ($r<0.2R$) from all
       simulations. The gray dashed line indicates 60 kG, which is the
       saturated value of the core dynamo from MHDr2.}
    \label{fig:Brms-core}
\end{figure}

All of the current simulations host a strong core dynamo. However, in
some cases the dynamo is
enhanced by the fossil field while in others it remains mostly
unaffected (see
Table~\ref{table-results}). More specifically, runs with Dipole A have
essentially the same $\bm{B_\mathrm{rms}}$ as the run with
no added dipole (MHDr2), irrespective of the inclination angle 
(see upper panel of Figure~\ref{fig:Brms-core}). On the other hand,
runs with Dipoles B and B* have stronger core magnetic fields, in some cases thrice that of
MHDr2. Interestingly, in the latter two sets there is a relation
between the strength of the magnetic fields and the obliquity angle
$\beta$.

\subsubsection{Enhanced dynamos \label{enhanced-dynamos}}

Figure~\ref{fig:Brms-core} shows that runs with Dipoles B and B* start
with the same core magnetic field amplitudes, and that only the cases
with $\beta=90\degr$ decay to the same level as the core dynamo
without any additional dipole field (the gray dashed line). The
aligned cases
(DipB, DipB*) and the rest of inclinations (DipBt, DipBt2, DipBt*,
DipBt2*, DipBt3*,DipBt4*) saturate in higher values than those of run
MHDr2, suggesting enhanced core dynamo action. The saturation levels
of these runs are similar although temporal averages (fourth column of
Table~\ref{table-results}) show that there is an inverse relation
between $\beta$ and $B_\mathrm{rms}$. The strongest rms magnetic
fields
for both averages, are obtained when the
dipole is aligned with the rotation axis. In the misaligned cases the
field decreases with increasing $\beta$. The only
exception to this trend is DipBt*, which has the same averaged
flow velocity and rms magnetic fields as DipB*. Interestingly,
inside the core, DipBt* has even a stronger toroidal component than
DipB*,
while in the rest of runs the inverse relation applies for both
poloidal and toroidal components. Furthermore, weaker magnetic fields
lead to weaker quenching of flows. It is worth mentioning that
although we
only added a strong poloidal component in the simulations, the
toroidal component of the core dynamo also increases in the enhanced
cases. The cases with horizontal dipoles decayed almost
completely. These runs will be analyzed separately in
Section~\ref{unaffected}.

\begin{figure*}[h!]
    \centering
    \includegraphics[scale=0.43]{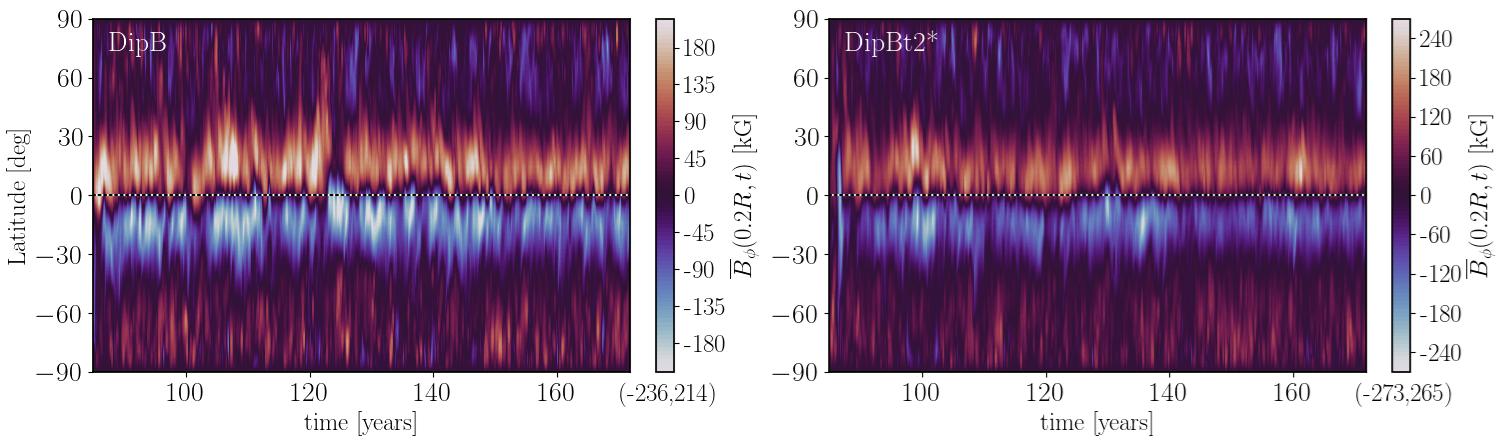}
    \caption{Time-latitude diagrams of the azimuthally averaged
      toroidal magnetic field $\overline{B}_\phi (r=0.2R,\theta,t)$ of
      runs DipB and DipBt2*. The run label is indicated in the upper left
      corner of each panel.}
    \label{fig:B-enhanced}
\end{figure*}

The azimuthally averaged toroidal magnetic fields at $r=0.2R$ as a
function of time and latitude for representative enhanced cases are
shown in Fig.~\ref{fig:B-enhanced}. The initially cyclic and
hemispheric
solutions \citepalias[see Fig.~2 of][]{Hidalgo2024} are replaced by
quasi-stationary solutions. All the simulations with an enhanced core
dynamo end up in the same configuration, irrespective of the obliquity
angle. Quasi-stationary dynamo solutions appear in simulations either
with low Coriolis numbers
\citep[e.g.][]{Kapyla-2013,2018ApJ...863...35S,Brun_et_al_2022_ApJ_926_21,
  ortizrodriguez-2023}, or in the strong field branch at rapid
rotation
\citep[e.g.][]{2006GeoJI.166...97C,Gastine-2012,YCMGRPW15}. The
current simulations land in the latter due to the strong added dipole
fields. The obtained configuration
resembles a dipole, with a positive polarity in latitudes from
$0^\circ$ to $30^\circ$ and negative polarity from $0^\circ$ to
$-30^\circ$. Beyond these latitudes, the polarity reverses, and the
magnetic fields are weaker than in the latitudes close to the equator. As
the added dipole field is purely poloidal, the toroidal fields likely
come from a dynamo process in the convective core.

\begin{figure}[t!]
    \centering
    \includegraphics[scale=0.35]{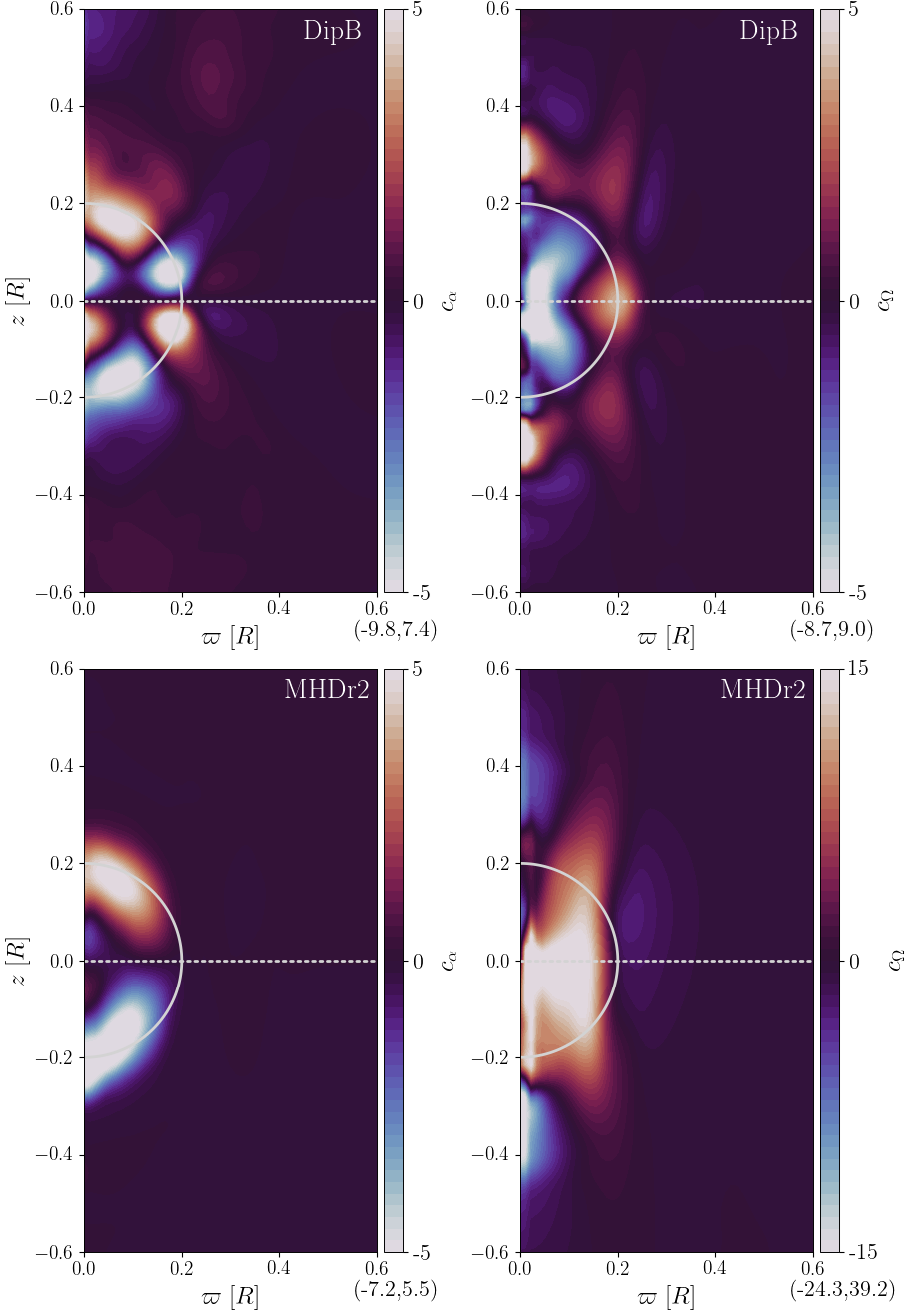}
     \caption{Dynamo parameters $c_\alpha = \alpha \Delta r/
       \eta_\mathrm{turb}$ and $c_\Omega = \partial
       \overline{\Omega}/\partial r(\Delta r)^3/\eta_\mathrm{turb}$
       from DipB (\textit{upper panels}) and MHDr2 (\textit{bottom
         panels}). All the panels are clipped and cropped (to $0.6R$)
       for better legibility and comparison.}
    \label{fig:DipBt2-params}
\end{figure}

To understand the origin of these dynamos, we make use of dynamo
numbers based on mean-field dynamo theory
\citep{KR80,Brandenburg-2005}. To quantify
the $\alpha$ and $\Omega$ effects, it is useful to define
\citep{Kapyla-2013}
\begin{equation}
    c_\alpha = \frac{\alpha \Delta r}{\eta_\mathrm{turb}},\hspace*{0.5cm} c_\Omega = \frac{\partial \overline{\Omega}/\partial r (\Delta r)^3}{\eta_\mathrm{turb}}, \label{dyn-param}
\end{equation}
where $\eta_\mathrm{turb}=\tau u_\mathrm{rms}^2 /3$ is an estimate of
the turbulent
diffusivity, $\tau = \Delta r/u_\mathrm{rms}$ is the convective
turnover time, and $\overline{\Omega}$ is the time- and azimuthally
averaged rotation rate. The non-linear $\alpha$ effect is proportional
to the difference between the fluctuating kinetic and magnetic
helicities, given by \citep{PFL76}
\begin{equation}
    \alpha = - \frac{\tau}{3}\left[(\overline{\bm{U} \bm{\cdot} \bm{\omega}} - \overline{\bm{U}} \bm{\cdot} \overline{\bm{\omega}})-(\overline{\bm{J} \bm{\cdot} \bm{B}} - \overline{\bm{J}} \bm{\cdot} \overline{\bm{B}})/\overline{\rho}\right],
\end{equation}
where $\bm{\omega} = \bm{\nabla}\times \bm{U}$ is the vorticity. The dynamo
parameters (\ref{dyn-param}) from DipB and MHDr2 are shown in
Figure~\ref{fig:DipBt2-params}. 
The profiles and amplitudes of
$c_\alpha$ in the current simulations
are similar to those of runs without a fossil field
(see left panels of Fig.~\ref{fig:DipBt2-params}). 
However, the enhanced cases show zones of strong $\alpha$ effect 
with a different sign close
to the equator. 
These are also the locations where the magnetic field is the
strongest (see Fig.~\ref{fig:B-enhanced}). The profile of $c_\alpha$
is antisymmetric with respect to the equator. The
profiles of $c_\Omega$ are of opposite sign in comparison to those of
runs
in \citetalias{Hidalgo2024}. This is a consequence of the
anti-solar differential rotation of the core (see
Section~\ref{DR}). The amplitudes of $c_\Omega$ are also
much weaker than in \citetalias{Hidalgo2024}. 
This is clearly seen in the right panels of 
Fig.~\ref{fig:DipBt2-params}.
The
inclusion of a strong imposed field quenches the
differential rotation, reducing the shear, which is translated in
lower values of $c_\Omega$. Table~\ref{table-energy} shows that
the energy in differential rotation in the core and in the radiative
envelope
is weaker than in MHDr2. Furthermore, in the enhanced dynamos,
the poloidal energy is stronger than the toroidal energy, being
typically $E_\mathrm{mag}^\mathrm{pol} \approx 3
E_\mathrm{mag}^\mathrm{tor}$. 

We interpreted the core dynamos from \citetalias{Hidalgo2024} as
$\alpha\Omega$ or $\alpha^2\Omega$ dynamos, primarily because the
signs of $\alpha$, $\partial_r \overline{\Omega}$ and the propagation
direction of the dynamo waves, and secondarily, because $c_\Omega \gg
c_\alpha$ and the dominance of toroidal over poloidal fieds. In the
enhanced
dynamos, the lower values of $c_\Omega$, weaker than $c_\alpha$ in
some runs, and the ratio
$E_\mathrm{mag}^\mathrm{pol}/E_\mathrm{mag}^\mathrm{tor} \approx 3$
suggest an $\alpha^2$ dynamo. These dynamos rely on the
$\alpha$ effect, and no shear from differential rotation is directly
needed. Their initial formulation suggested only non-oscillatory
dynamos, for example, the axisymmetric models by
\cite{1969AN....291...49S},
which fits with the quasi-stationary solutions of the axisymmetric
toroidal magnetic field found in our simulations. However, oscillatory
solutions have also been found in numerical simulations
\citep{1987AN....308...89B, 2010ApJ...719L...1M, Kapyla-2013, 2014ApJ...794L...6M} and
analytical work \citep{1987AN....308..101R, 2017A&A...598A.117B}.

\begin{figure}[t!]
    \centering
    \includegraphics[width=\hsize]{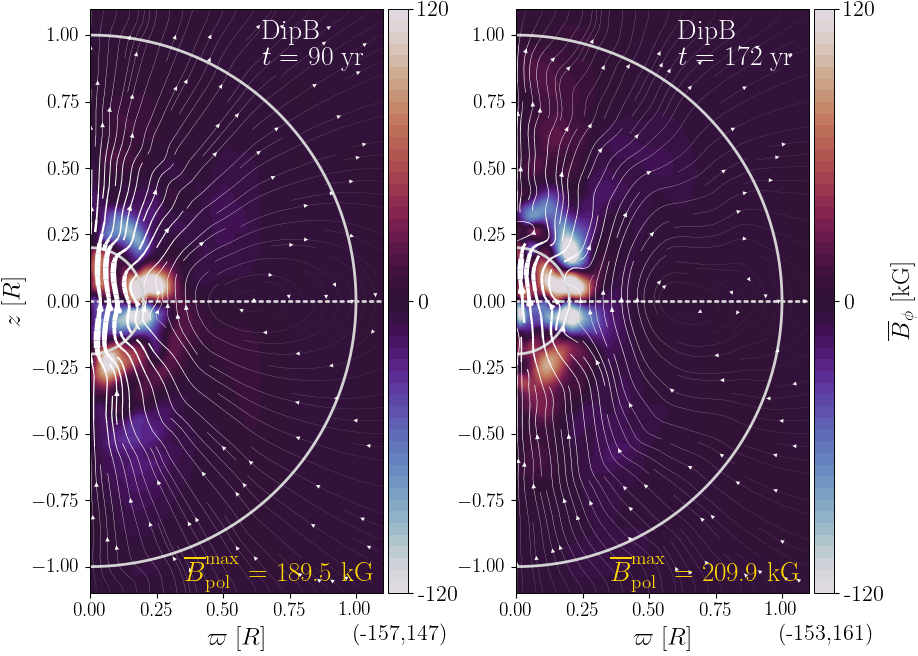}
     \caption{Azimuthally averaged toroidal magnetic field
       $\overline{B}_\phi(\varpi,z)$ of DipB. The poloidal magnetic
       field is represented with arrows, where the width is
       proportional to the strength of the field. The values of
       $\overline{B}_\phi$ are clipped and the maximum and minimum
       values $(\overline{B}_\phi^\mathrm{min},
       \overline{B}_\phi^\mathrm{max})$ are indicated below the
       colorbar. The dashed line at $z = 0$ represents the equator. The complete temporal evolution is available as an online movie.}
    \label{fig:DipB-comp}
\end{figure}

The meridional distribution of the azimuthally averaged magnetic field
of DipB is shown in Figure~\ref{fig:DipB-comp}. In an early stage of
the simulation, five years ($\sim 240 \tau_\mathrm{ff}$) after the
inclusion of the fossil field, the core dynamo already exhibits the
described quasi-stationary behavior (see left panel). Furthermore,
some of the toroidal magnetic fields
are seen outside the convective core,
at the base of the radiative envelope. The poloidal magnetic field
follows the structure of Dipole B outside the core, while within the
field is nearly vertical like in Dipole A. The
magnetic field in the radiative zone is different than that generated
by the core dynamo without an added dipole \citepalias[see Fig.~4
  of][]{Hidalgo2024}. Initially, the magnetic field 
spread close to the rotation axis
  from $z
  \approx 0.3R$ to $r \approx 0.5R$, which is the location of the
  magnetic diffusivity jump. In \citetalias{Hidalgo2024} we mention
that this magnetic field is most likely transported by vertical flows
(see, for example, the rightmost panel of Fig.~\ref{fig:MHDr2-B}),
after which it evolves on
long timescales compared to the period of the core dynamo due to the
low diffusivities there. In the current simulations, there is much
less magnetic fields beyond the radial jump of diffusivities
($r\approx
0.35R$), indicating that in runs with the enhanced core dynamo the
flows are not transporting magnetic field from the core to this zone
as efficiently. The right panel of Figure~\ref{fig:DipB-comp} displays
the
magnetic field of DipB 82 years ($\sim 3993
\tau_\mathrm{ff}$) after that in left panel. The quasi-stationary
nature of the core dynamo remains
unaffected. Furthermore, the dipolar structure present in most of the
radiative envelope and at the stellar surface remains essentially
unchanged, indicating that the enhanced core dynamo influences the
surface poloidal magnetic field of the star only weakly (see
Section~\ref{surface-fields}).

\begin{table*}[t!]
\centering
\caption{Kinetic and magnetic energies.}
\begin{tabular}{lccccccc}
\hline\hline\noalign{\smallskip}
Run & $E_\mathrm{kin}~[10^{33}\mathrm{J}]$  & $E_\mathrm{kin}^\mathrm{DR}/E_\mathrm{kin}$ & $E_\mathrm{kin}^\mathrm{MC}/E_\mathrm{kin}$ & $E_\mathrm{mag}~[10^{33}\mathrm{J}]$ & $E_\mathrm{mag}^\mathrm{tor}/E_\mathrm{mag}$ & $E_\mathrm{mag}^\mathrm{pol}/E_\mathrm{mag}$ & $E_\mathrm{mag}/E_\mathrm{kin}$\\
\hline\noalign{\smallskip}
Convective core \\
\hline
MHDr2 & 8.88 & 0.346 & 0.026 & 1.74 & 0.140 & 0.097 & 0.196 \\
DipA & 8.60 & 0.321 & 0.025 & 1.72 & 0.109 & 0.143 & 0.200 \\
DipAt & 8.65 & 0.336 & 0.024 & 1.64 & 0.114 & 0.137 & 0.190 \\
DipAt2 & 8.78 & 0.343 & 0.024 & 1.72 & 0.130 & 0.119 & 0.196 \\
DipAh & 9.48 & 0.369 & 0.028 & 1.66 & 0.137 & 0.096 & 0.176 \\
DipB* & 2.70 & 0.082 & 0.029 & 13.77 & 0.102 & 0.342 & 5.105 \\
DipBt* & 2.63 & 0.108 & 0.027 & 13.77 & 0.147 & 0.303 & 5.229 \\
DipBt2* & 3.24 & 0.055 & 0.021 & 9.34 & 0.104 & 0.290 & 2.879 \\
DipBt3* & 4.20 & 0.045 & 0.014 & 6.36 & 0.077 & 0.212 & 1.514 \\ 
DipBt4* & 4.57 & 0.035 & 0.014 & 5.19 & 0.071 & 0.173 & 1.135 \\
DipBh* & 9.77 & 0.358 & 0.025 & 1.35 & 0.108 & 0.089 & 0.138 \\
DipB & 2.92 & 0.063 & 0.024 & 11.40 & 0.115 & 0.324 & 3.906 \\
DipBt & 3.11 & 0.055 & 0.024 & 10.04 & 0.115 & 0.310 & 3.229 \\
DipBt2 & 3.72 & 0.050 & 0.017 & 7.78 & 0.079 & 0.241 & 2.092 \\
DipBh & 9.72 & 0.370 & 0.027 & 1.48 & 0.134 & 0.104 & 0.152 \\
\hline
Radiative envelope \\
\hline
MHDr2 & 32.37 & 0.470 & 0.006 & 16.08 & 0.378 & 0.041 & 0.497 \\
DipA & 25.54 & 0.355 & 0.007 & 52.14 & 0.460 & 0.119 & 2.042 \\
DipAt & 25.65 & 0.360 & 0.007 & 46.38 & 0.427 & 0.074 & 1.808 \\
DipAt2 & 27.28 & 0.399 & 0.006 & 37.99 & 0.376 & 0.092 & 1.393 \\
DipAh & 31.47 & 0.453 & 0.007 & 21.60 & 0.303 & 0.035 & 0.686 \\
DipB* & 9.00 & 0.129 & 0.005 & 89.82 & 0.165 & 0.220 & 9.983 \\
DipBt* & 9.47 & 0.175 & 0.005 & 91.13 & 0.211 & 0.170 & 9.628 \\
DipBt2* & 8.89 & 0.088 & 0.003 & 61.65 & 0.175 & 0.114 & 6.932 \\ 
DipBt3* & 10.49 & 0.061 & 0.003 & 46.78 & 0.108 & 0.047 & 4.460 \\
DipBt4* & 11.75 & 0.046 & 0.003 & 41.71 & 0.109 & 0.036 & 3.550 \\
DipBh* & 27.39 & 0.368 & 0.008 & 32.96 & 0.163 & 0.024 & 1.203 \\
DipB & 8.63 & 0.107 & 0.004 & 70.46 & 0.224 & 0.151 & 8.169 \\
DipBt & 8.74 & 0.088 & 0.004 & 68.53 & 0.184 & 0.125 & 7.840 \\
DipBt2 & 9.24 & 0.070 & 0.003 & 52.21 & 0.145 & 0.074 & 5.651 \\
DipBh & 29.75 & 0.412 & 0.007 & 32.13 & 0.257 & 0.037 & 1.080 \\
\hline
\end{tabular}
\label{table-energy}
\tablefoot{The total kinetic energy is $E_\mathrm{kin} = \frac{1}{2}
  \int \rho \bm{U}^2 dV$. The energies for the differential rotation
  (DR) and meridional circulation (MC), are
  $E_\mathrm{kin}^\mathrm{DR} = \frac{1}{2} \int \rho
  \overline{U}_\phi^2 dV$ and $E_\mathrm{kin}^\mathrm{MC} =
  \frac{1}{2} \int \rho (\overline{U}_\varpi^2 + \overline{U}_z^2)
  dV$, respectively. The total magnetic energy is $E_\mathrm{mag} =
  \frac{1}{2\mu_0} \int \bm{B}^2 dV$, and the toroidal and poloidal
  magnetic energies are $E_\mathrm{mag}^\mathrm{tor} =
  \frac{1}{2\mu_0} \int \overline{B}_\phi^2 dV$ and
  $E_\mathrm{mag}^\mathrm{pol} = \frac{1}{2\mu_0} \int
  (\overline{B}_\varpi^2 + \overline{B}_z^2) dV$, respectively. The
  energies are averaged over time and integrated over the convective core and the radiative envelope. The total kinetic and magnetic
  energies are given in units of $10^{33}\mathrm{J}$.}
\end{table*}

\begin{figure*}[h!]
    \centering
    \includegraphics[scale=0.43]{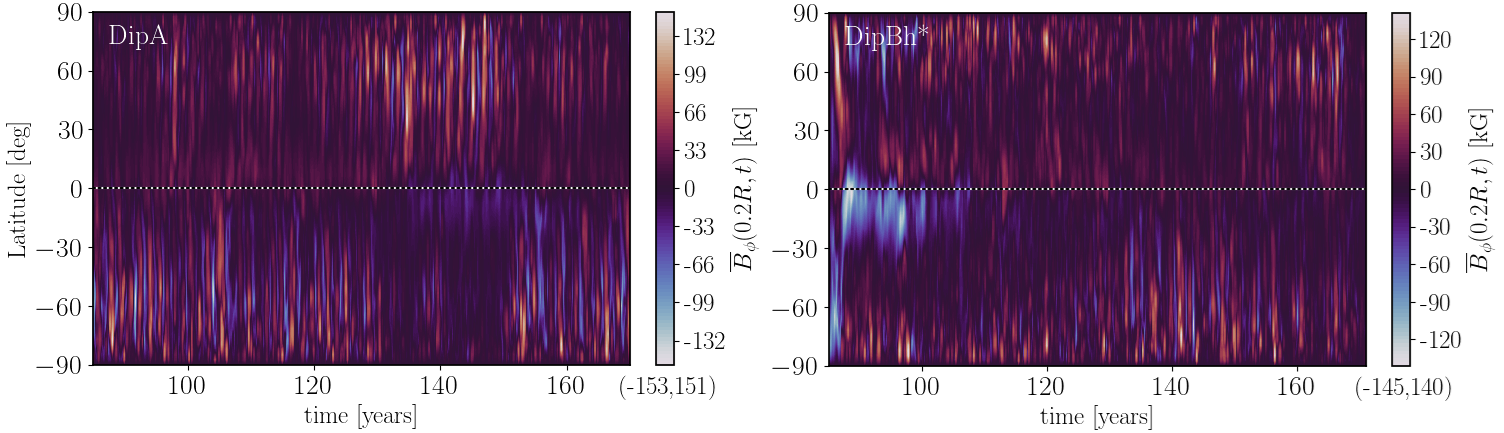}
    \caption{Same as Figure 4, but for runs without an enhanced dynamo.}
    \label{fig:B-unaffected}
\end{figure*}

Finally, these runs have superequipartition magnetic fields. More
specifically, DipB* and DipBt* have
$E_\mathrm{mag}/E_\mathrm{kin} \approx 5$ in the convective core, and
$E_\mathrm{mag}/E_\mathrm{kin} \approx 10$ in the
radiative envelope. The core dynamo of the fast rotators from
\cite{Augustson-2016} achieve similar regimes as those in our
simulations, for example, their fastest rotator (M16) reaches
$E_\mathrm{mag}/E_\mathrm{kin} = 5.02$. While still in
superequipartition, it is unclear how $E_\mathrm{mag}/E_\mathrm{kin}$
changes in the simulations of
\cite{Featherstone-2009} with an added purely poloidal
field. Nevertheless, the inclusion of a mixed field (with poloidal and
toroidal components) increases the ratio in the core dynamo to
$E_\mathrm{mag}/E_\mathrm{kin} \approx 10$. This ratio is comparable
to what we find in the radiative envelope of our runs but larger than
in our core dynamos. In the mixed case of \cite{Featherstone-2009},
$E_\mathrm{mag}/E_\mathrm{kin} \approx 5.18$ in the radiative envelope
(see their Table 1), which is less than in our simulations. This might
be because they consider only a fraction of the radiative envelope. In
runs with Dipole
B, the enhanced dynamos are also in the superequipartition regime, but
the ratios are somewhat smaller than those from runs with Dipole B*,
both in the convective core and in the radiative zone.

\subsubsection{The unaffected cases \label{unaffected}}

The azimuthally averaged toroidal magnetic fields at $r=0.2R$ of
representative 
simulations without an enhanced dynamo are shown in
Figure~\ref{fig:B-unaffected}. None of the core dynamos
from runs with Dipole A got enhanced, and the obliquity angle of the
dipole has no effect in the magnetic field configuration. 
This is likely a consequence of
the strength of the dipole. While all the
configurations have a surface dipolar amplitude of 6 kG, Dipole A
keeps this amplitude constant inside
the star, and therefore, $B_\mathrm{rms}^\mathrm{pol}$ inside the core
is
$\sim 6$ kG. This is much weaker than the fields from Dipoles B and B*
(see Section~\ref{dipolar-fossil}) or that from the core dynamo of
MHDr2
($\sim 19$ kG). Furthermore,
their magnetic cycle period is not affected either,
being essentially the same as that 
from MHDr2 (see the last
column of Table~\ref{table-results}). 
Figure~\ref{fig:B-unaffected} shows that
the core dynamo from DipA remains cyclic and hemispheric, occasionally
changing its active hemisphere like run 
MHDr2* of \citetalias{Hidalgo2024} and in the fully convective model
of \cite{brown-2020}.
All the runs with Dipole A show at least one dynamo migration except
DipAh. One difference between these runs and MHDr2 is that
close to the equator a weak structure similar to the quasi-stationary
solutions of the enhanced dynamos is visible, except in DipAh. For
example, this structure can be spotted in DipA where it is clearest
between $\sim 130$ and $\sim 150$ years.

The right panel of Figure~\ref{fig:B-unaffected} display the
toroidal magnetic field of the horizontal case ($\beta = 90^\circ$)
with Dipole B*. In this case, the core dynamo
exhibits the quasi-stationary solution described in
Section~\ref{enhanced-dynamos} right after the inclusion of the
dipole field. However, the solution is unstable and it decays after a
few years, returning to the original hemispherical configuration of
the core dynamo. The same happens in DipBh.
Interestingly, this fast decay happens exclusively in
the horizontal cases, whereas even the cases with $\beta = 80^\circ$
and $\beta = 85^\circ$ have a stable enhanced dynamo with at least
twice the strength of the horizontal cases (see
Table~\ref{table-results}).

\subsection{Stability and surface magnetic fields \label{surface-fields}}

\begin{figure*}[t!] 
    \centering
    \includegraphics[scale=0.54]{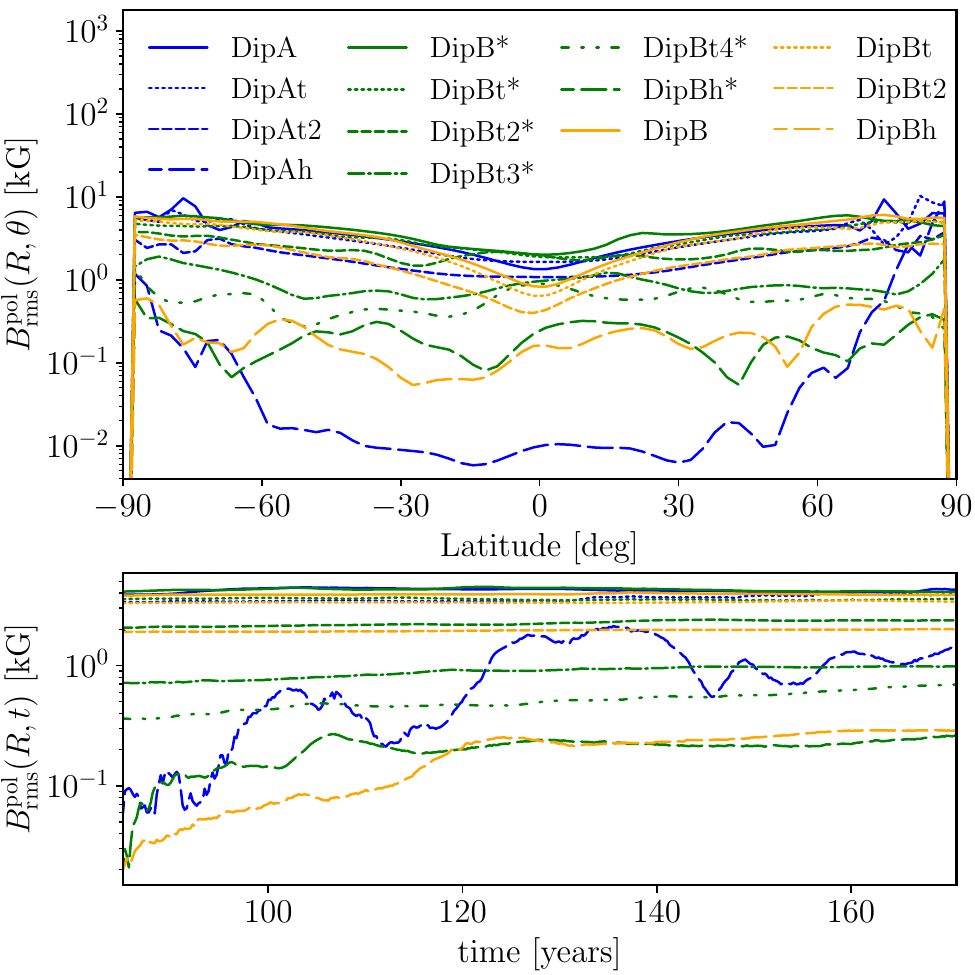}
    \includegraphics[scale=0.54]{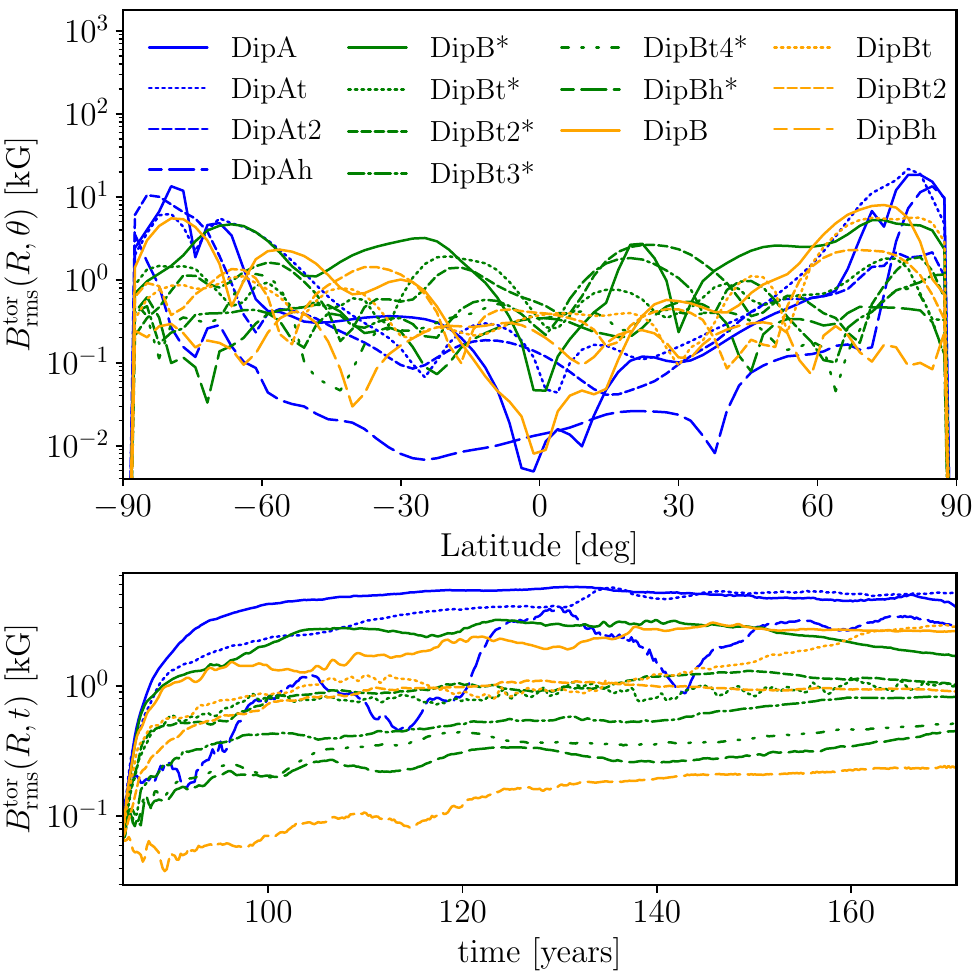}
     \caption{Distribution and temporal evolution of the rms magnetic
       field at the stellar surface ($r=R$) of all the
       simulations. \textit{Upper panels:} Poloidal (left) and
       toroidal (right)
       components of the azimuthally and temporally averaged
       rms-field as a function of latitude. \textit{Bottom
         panels:} Temporal evolution of the horizontally ($\phi
       \theta$) averaged poloidal (left) and toroidal (right)
       rms-field.}
    \label{fig:Brms-surf}
\end{figure*}

\begin{figure*}[t!] 
    \centering
    \includegraphics[scale=0.49]{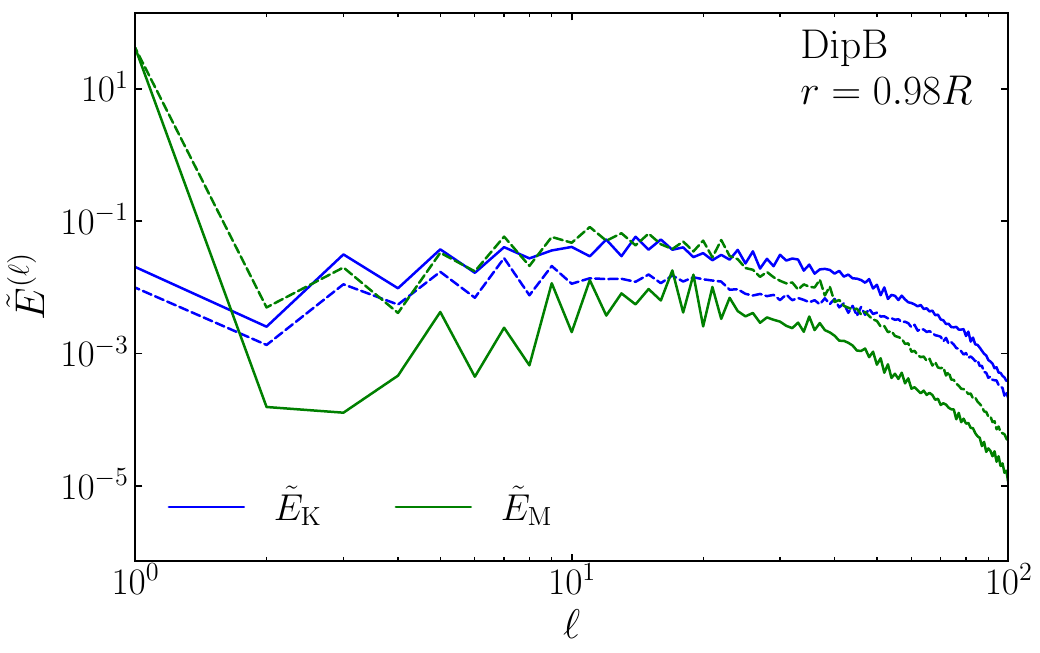}
     \includegraphics[scale=0.49]{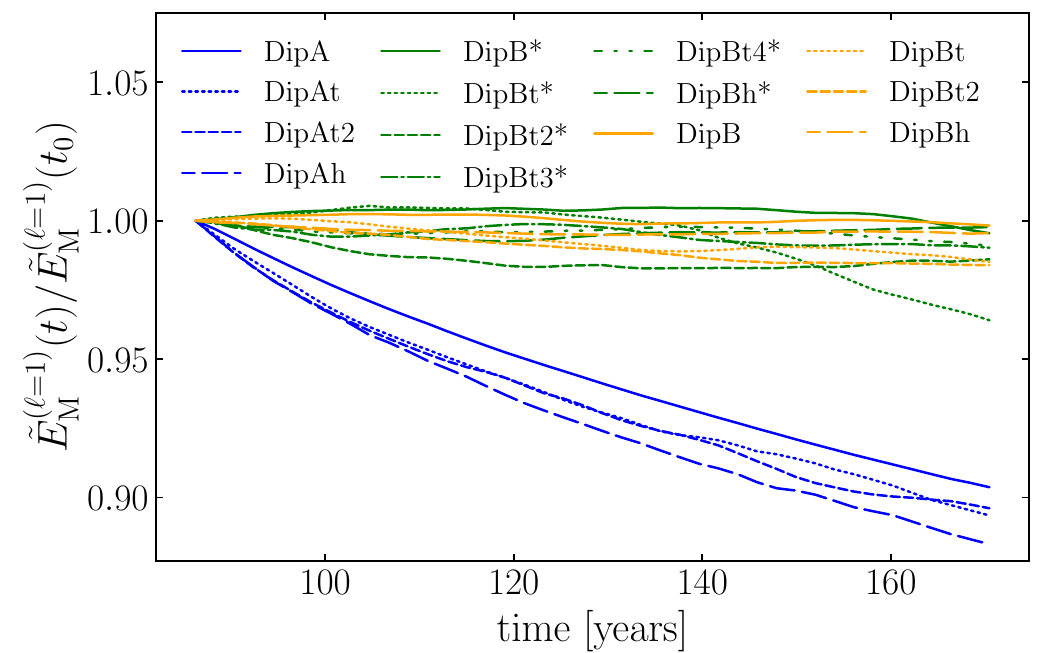}
     \caption{\textit{Left panel:} Normalized power spectra of the
       velocity $\tilde{E}_\mathrm{K}^{(\ell)} =
       E_\mathrm{K}^{(\ell)}/ \sum_l E_\mathrm{K}^{(\ell)}$ and
       magnetic fields $\tilde{E}_\mathrm{M}^{(\ell)} =
       E_\mathrm{M}^{(\ell)}/ \sum_l E_\mathrm{K}^{(\ell)}$ from DipB
       at $r=0.98R$. The solid lines show the power spectra from early
       times (first 15 years) of the simulation, and the dashed lines
       the late times (last 15 years). \textit{Right panel:} Temporal
       evolution of the dipolar contribution ($\ell=1$) of the power
       spectra of the magnetic field $\tilde{E}_\mathrm{M}$ from all
       the simulations. The values were normalized by
       $\tilde{E}_\mathrm{M}^{(1)}$ at the starting point of each
       simulation (at $t_0 = 85$ years), for a better comparison.}
    \label{fig:PS-surf}
\end{figure*}

In principle, a purely poloidal magnetic field whose field lines are
closed inside the star, like Dipole B and Dipole B*, is unstable
against adiabatic perturbations \citep{Markey-1973, Wright-1973,
  Markey-1974}. Similarly, a poloidal field with none of the field
lines closed inside the star, like Dipole A, is unstable as well
\citep{Flowers1977}. However, these analytical results are highly
idealized, as dissipative effects and rotation are not
considered. Furthermore, we are not testing the stability of a purely
poloidal magnetic field. In MHDr2, 
 $37.8 \%$ of the magnetic energy
inside the radiative envelope is toroidal, and as described in
Section~\ref{enhanced-dynamos}, the added poloidal field can also
induce a toroidal component close to the core. Therefore, the
radiative envelopes in our simulations have both components in all the
cases, which as suggested by analytical estimations
\citep{1956ApJ...123..498P, Wright-1973}, and numerical results
\citep{2004Natur.431..819B, Braithwaite-2006}, seems to be a necessary
condition to achieve stability. Furthermore, the stability condition
reported by \cite{2009MNRAS.397..763B} suggests $10^{-5} \lesssim
E_\mathrm{mag}^\mathrm{pol}/E_\mathrm{mag} \lesssim 0.8$ for a MS
A-type star, which is fulfilled inside the radiative envelope in all
of
our simulations (see column 7 of Table~\ref{table-energy}).

The upper panels of Figure~\ref{fig:Brms-surf} show the latitudinal
distribution of the poloidal (left) and toroidal (right) components of
azimuthally and temporally averaged rms-field at the stellar
surface. In the cases with $\beta\neq90\degr$, prominent poloidal
fields are seen in the full range of latitudes with slightly weaker
amplitudes around the equator. Interestingly, a toroidal component
with
roughly the same amplitude is also present in these runs.
Distribution of $B_\phi$ at the stellar surface is less coherent than
the poloidal field, with significantly weaker values at the equator and
maxima close to the poles. This is similar to the runs of
\citetalias{Hidalgo2024} (see their Fig.~6), although there the fields
are much weaker ($B_\mathrm{rms} \sim 10^{-5}$ kG) apart from the
poles. Temporal evolution of the horizontally averaged
$B_\mathrm{rms}^\mathrm{pol}$ is in the bottom left panel of 
Fig.~\ref{fig:Brms-surf}. This
component remains essentially unaffected in the runs with $\beta \neq
0$, while in the cases with $\beta = 0$ it reaches saturation after
$\sim 120$ years in runs DipBh and DipBh*. The case of DipAh is
different,
because most of its activity is located in the poles, so the observed
surface field is most likely a consequence of the core dynamo and the
vertical flows, same as in Fig.~6 of \citetalias[][]{Hidalgo2024}, and
not the fossil field that we are
adding. From the bottom right panel it is visible that the toroidal
component from all the runs is a consequence of the imposed fossil
field, as it starts increasing right after its inclusion. In the runs
with $\beta \neq 0$ it grows until it roughly reaches the strength of
the added dipole field. The origin of this toroidal component is
not clear, but dynamo action in layers close to the
surface seems highly unlikely, as the surface has almost no shear (see
Section~\ref{DR}) and both dynamo parameters are close to
zero. Therefore, this toroidal field is most likely transported from the
enhanced core dynamo, or from the bottom radiative envelope. After the
inclusion of Dipoles B and B*, the advection timescales
increased from 100 yrs (see \citetalias{Hidalgo2024}) to 1000 and 640
yrs, respectively. However, turbulent diffusion timescales are around
11 yrs in
regions closer to the axis of rotation, and around 50 yrs in other
latitudes. Therefore, the toroidal magnetic field is most likely
transported to the surface due to turbulent diffusion.

The normalized power spectra of the velocity and magnetic field from a
spherical harmonic decomposition \citep[see][]{KR80}
of DipB is shown in the left panel of
Fig.~\ref{fig:PS-surf}. The solid lines correspond to the power
spectra of the first 15 years of the simulation, and the dashed lines
of the last 15 years. It is clear that in both periods the magnetic
power spectrum has its peak in the dipolar term with $\ell=1$. Moreover,
during the early times, $99.54 \%$ of the magnetic energy is
in the dipole component. At late times, this reduces to $97.27\%$, but
this is not because the dipole amplitude decreases, but due to an
increase in the contribution of smaller scales ($\ell > 1$).
The time evolution of the dipole amplitude ($\ell = 1$), normalized by
its initial value, is shown in the right panel of
Fig.~\ref{fig:PS-surf}. In the cases with Dipole B and B*, the dipole
amplitude is essentially constant over time, with the exception of
DipBt* where the dipole decays more than in the other
runs. Nevertheless, the fraction of the energy in the dipole changes
from $99.33\%$ at early to $94.19\%$ at late times. In the runs with
Dipole A, the
dipole decays faster than in the rest of simulations. For example, in
DipAt the dipole contains $98.98\%$ of the energy at early times, but
at the late times this fraction drops to $50.28\%$.
This suggests that while Dipole B can enhance the dynamo and possibly
land it on the strong field branch, Dipole A is not conducive to this.

\subsection{Differential rotation \label{DR}}

\begin{table*}[t!]
\centering
\caption{Differential rotation parameters and the maximum meridional flow $\overline{U}_\mathrm{mer}^\mathrm{max}$ from all the simulations.
}
\begin{tabular}{lccccccc}
\hline\hline\noalign{\smallskip}
Run & $\Delta_\Omega^{(r)}$ & $\Delta_\Omega^{(\overline{\theta})}(60^\circ)$ & $\Delta_\Omega^{(\overline{\theta})}(75^\circ)$ & $\Delta_\Omega^{\mathrm{CZ}(r)}$ & $\Delta_\Omega^{\mathrm{CZ}(\overline{\theta})}(60^\circ)$ & $\Delta_\Omega^{\mathrm{CZ}(\overline{\theta})}(75^\circ)$ & $\overline{U}_\mathrm{mer}^\mathrm{max}~\mathrm{[m/s]}$ \\
\hline\noalign{\smallskip}
MHDr2  &  0.0544  &  $-3.683\cdot 10^{-5}$  &  $-5.329\cdot 10^{-4}$ &  0.1201  &  0.0714  &  0.1196 & 6.0 \\ \hline
DipA  &  0.0515  &  $1.223\cdot 10^{-6}$  &  $-7.563\cdot 10^{-5}$ &  0.1056  &  0.0621  &  0.1059 & 3.6 \\
DipAt  &  0.0518  &  $-1.236\cdot 10^{-5}$  &  $-1.256\cdot 10^{-5}$ &  0.1077  &  0.0626  &  0.1067 & 2.1 \\
DipAt2  &  0.0508  &  $5.012\cdot 10^{-6}$  &  $3.177\cdot 10^{-5}$ &  0.1056  &  0.0634  &  0.1055 & 1.8 \\
DipAh  &  0.0533  &  $1.257\cdot 10^{-5}$  &  $-2.068\cdot 10^{-5}$ &  0.1117  &  0.0675  &  0.1130 & 5.2 \\ \hline
DipB*  &  0.0144  &  $9.323\cdot 10^{-6}$  &  $6.763\cdot 10^{-5}$ &  0.0107  &  0.0039  &  0.0029 & 2.3 \\
DipBt*  &  0.0153  &  $1.255\cdot 10^{-5}$  &  $2.354\cdot 10^{-5}$ &  0.0146  &  0.0036  &  0.0025 & 2.2 \\
DipBt2*  &  0.0104  &  $1.394\cdot 10^{-5}$  &  $5.852\cdot 10^{-5}$ &  0.0000  &  -0.0085  &  -0.0151  & 2.0 \\
DipBt3*  &  0.0065  &  $5.966\cdot 10^{-6}$  &  $3.665\cdot 10^{-5}$ &  -0.0069  &  -0.0134  &  -0.0228  & 1.9 \\ 
DipBt4*  &  0.0102  &  $1.362\cdot 10^{-5}$  &  $2.461\cdot 10^{-5}$ &  0.0043  &  -0.0056  &  -0.0136 & 2.8 \\
DipBh*  &  0.0378  &  $4.459\cdot 10^{-6}$  &  $9.171\cdot 10^{-6}$ &  0.0782  &  0.0382  &  0.0745 & 4.1 \\ \hline
DipB  &  0.0128  &  $1.438\cdot 10^{-7}$  &  $1.029\cdot 10^{-4}$ &  0.0063  &  -0.0037  &  -0.0080 & 1.9 \\
DipBt  &  0.0117  &  $7.885\cdot 10^{-6}$  &  $7.158\cdot 10^{-5}$ &  0.0045  &  -0.0060  &  -0.0102 & 1.6 \\
DipBt2  &  0.0061  &  $1.375\cdot 10^{-5}$  &  $2.895\cdot 10^{-5}$ &  -0.0099  &  -0.0150  &  -0.0257 & 3.3 \\
DipBh  &  0.0485  &  $1.978\cdot 10^{-5}$  &  $1.097\cdot 10^{-6}$  &  0.1053  &  0.0563  &  0.1063   & 2.1 \\
\hline

\end{tabular}
\label{table-DR2}
\tablefoot{The differential rotation parameters follow Eqs.~(\ref{rot-param1}) and (\ref{rot-param2}).}
\end{table*}

The averaged rotation rate is given by
\begin{equation}
    \overline{\Omega}(\varpi,z) = \Omega_0 + \overline{U}_\phi(\varpi,z)/\varpi,
\end{equation}
where $\varpi = r\sin \theta$ is the cylindrical radius. Additionally, the averaged meridional flow is
\begin{equation}
    \overline{U}_\mathrm{mer}(\varpi,z) = (\overline{U}_\varpi,0,\overline{U}_z).
\end{equation}
To quantify the amplitude of the radial and latitudinal differential rotation, we use the following parameters \citep{Kapyla-2013}
\begin{align}
    \Delta_\Omega^{(r)} &= \frac{\overline{\Omega}(r_\mathrm{top}, \theta_\mathrm{eq}) - \overline{\Omega}(r_\mathrm{bot}, \theta_\mathrm{eq}) }{\overline{\Omega}(r_\mathrm{top}, \theta_\mathrm{eq})}, \label{rot-param1}\\ \Delta_\Omega^{(\overline{\theta})} &= \frac{\overline{\Omega}(r_\mathrm{top}, \theta_\mathrm{eq}) - \overline{\Omega}(r_\mathrm{top}, \overline{\theta}) }{\overline{\Omega}(r_\mathrm{top}, \theta_\mathrm{eq})}, \label{rot-param2}
\end{align}
where $r_\mathrm{top} = 0.9R$ and $r_\mathrm{bot} = 0.1R$ are the
radius near the surface and center of the star,
respectively. $\theta_\mathrm{eq}$ is the latitude at the equator, and
$\overline{\theta}$ is an average of $\overline{\Omega}$ between
latitudes $-\theta$ and $\theta$. Following \citetalias{Hidalgo2024},
we define $\Delta_\Omega^{\mathrm{CZ}(r)}$ and
$\Delta_\Omega^{\mathrm{CZ} (\overline{\theta})}$, which are the same
as Eqs. (\ref{rot-param1}) and (\ref{rot-param2}), but with
$r_\mathrm{top}=0.2R$ and $r_\mathrm{bot}=0.05R$, to study the
differential rotation of the core.

\begin{figure*}[h!]
    \centering
    \includegraphics[scale=0.38]{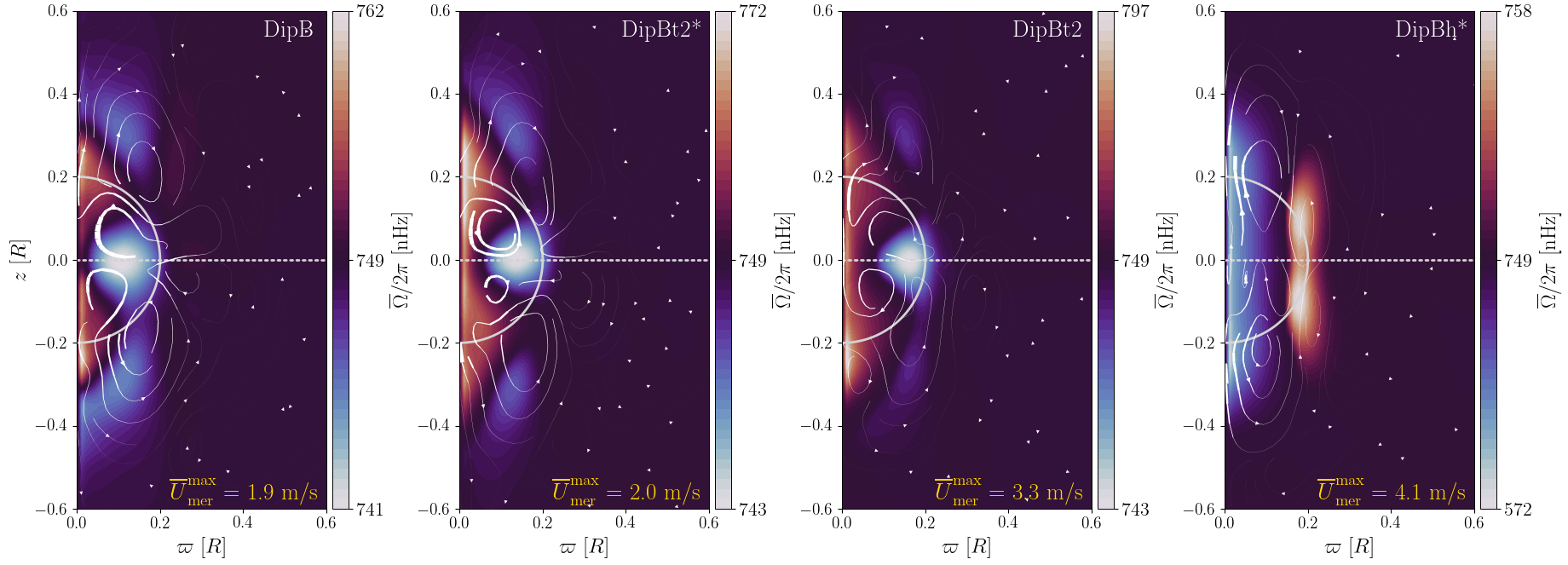}
    \caption{Profiles of the temporally and azimuthally averaged
      rotation rate $\overline{\Omega}(\varpi,z)$ of selected runs
      with Dipole B and Dipole B* (clipped at $r=0.6R$). The
      streamlines indicate the mass flux due to meridional
      circulation. The maximum meridional flow speed is indicated in
      the lower right side of the plot. The dashed line at $z=0$
      represents the equator.}
    \label{fig:pOm}
\end{figure*}

The differential rotation parameters and the maximum meridional speeds
are summarized in Table~\ref{table-DR2}. From columns 3 and 4 of
Table~\ref{table-DR2} we can conclude that MHDr2 already has an
almost-rigid rotation close to the surface \citepalias[see Fig.~8
  of][]{Hidalgo2024}. However, the inclusion of a fossil field,
irrespective of its type, reduces the amplitude of the differential
rotation almost completely with typical values being
$|\Delta_\Omega^{(\overline{\theta})}(60^\circ)| \sim 10^{-6}$ and
$|\Delta_\Omega^{(\overline{\theta})}(75^\circ)| \sim
10^{-5}$. A similar phenomena is observed in the convective core. In
runs with Dipole A, the parameters of the core maintain their sign
($\Delta_\Omega^{\mathrm{CZ}(r)},
\Delta_\Omega^{\mathrm{CZ}(\overline{\theta})} > 0$), which indicates
that the solar-like state from run MHDr2 is still present in these
simulations. However, the amplitude of these coefficients is
weaker than in MHDr2. In the cases with Dipole B* and
Dipole B, the differential rotation of the core is significantly
quenched. The solar-like state changes to an
anti-solar regime, where $\Delta_\Omega^{\mathrm{CZ}(r)},
\Delta_\Omega^{\mathrm{CZ}(\overline{\theta})} < 0$ typically. This is
also visible in the first three panels of Figure~\ref{fig:pOm}. The
initially retrograde column close to the axis of rotation (see
panel(d) of Fig.\ref{fig:MHDr2-B}) of MHDr2 changes to a weak prograde
column in the runs with the enhanced dynamo. A similar transition was
reported by \cite{Augustson-2016}, between their hydrodynamic and the
superequipartition magnetic cases (see their Fig.~7). Moreover, our
results are also in agreement with \cite{Featherstone-2009}, as the
prominent retrograde columns of the progenitor cases are replaced by
regions of slightly prograde rotation, and the radiative envelope is
almost rotating solidly after the inclusion of a poloidal fossil
field. Finally, from the last column of Table~\ref{table-DR2} we
conclude that there is no apparent relation between $\beta$ and the
maximum meridional speeds.

\section{Discussion and conclusions \label{conclusions}}

The interaction between the core dynamo of a $2.2M_\odot$ A-type star
and an imposed purely poloidal fossil field with a dipolar surface
strength of 6 kG, was explored using star-in-a-box MHD simulations
\citep{Dobler-2006,Kapyla-2021}. Furthermore, as the observed magnetic
fields from Ap stars are typically misaligned with their rotational
axis, obliquity angles $\beta$ between $0^\circ$ and $90^\circ$ were
tested. In the case where none of the magnetic field lines are closed
inside the star (Dipole A), the core dynamo does not seem to be
affected, irrespective of the chosen inclination. A possible
explanation
for this, is that the internal poloidal field of this initial
condition ($B_\mathrm{rms}^\mathrm{pol} \approx 6$ kG) is too weak to
excite a new dynamo mode. \cite{2004A&A...414.1065M} 
found that an external poloidal magnetic field of the same order as 
the poloidal component of the initial dynamo, is sufficient to 
significantly affect the dynamo. Following this argument, the dipole 
field needs to be
at
least $B_\mathrm{rms}^\mathrm{pol} = 19$ kG (see
Table~\ref{table-results}) to change the nature of the core dynamo in
our simulations, which is not achieved by Dipole A. In the cases where
the magnetic field lines of the imposed field are closed inside the
star (Dipole B and Dipole B*), $B_\mathrm{rms}^\mathrm{pol}$ in this
region is strong enough to affect the core dynamo. More specifically,
with most of the explored inclinations (from $0^\circ$ to $85^\circ$)
the initially cyclic and hemispheric solutions of the dynamo changes
to a mainly dipolar quasi-stationary configuration. The strength of
this enhanced dynamo seems to be inversely related to the obliquity
angle of the fossil field. The horizontal
cases ($\beta = 90^\circ$) start with the mentioned quasi-stationary
solution, but decay after a few years, proving to be
unstable. Inertia seems to play an important role in the geometry of
the dynamo, that is, dipolar solutions are expected when inertia is weak
\citep{2006GeoJI.166...97C, 2012E&PSL.333....9S}. This is in agreement
with the enhanced dynamos, where $\mathrm{Co} \approx 15$, and where
the addition of a strong dipole moves the simulation permanently to
the dipolar branch. When inertia
becomes relevant, multipolar solutions are expected. The runs in
the dipolar branch by \cite{Gastine-2012} have similar
dipole-dominated solutions like the enhanced cases, while runs in the
multipolar branch have a cyclic and hemispheric nature like MHDr2 (see
their Fig.12). The enhanced dynamos have superequipartition fields,
such that $E_\mathrm{mag}/E_\mathrm{kin}$ is between 1.1 and
5.2. These ratios are comparable to the fast rotators of
\cite{Augustson-2016}. However, we do not find fields as strong as
those reported by \cite{Featherstone-2009} in their mixed-case
($E_\mathrm{mag}/E_\mathrm{kin} \approx 10$). 

In the enhanced cases, the strong magnetic fields affect the profiles
of differential rotation, inducing nearly rigid rotation in
the radiative envelope, and changing the differential rotation in the
convective core from solar-like to anti-solar. A similar transition
was reported by \cite{Featherstone-2009} after the inclusion of a
fossil field, and by \cite{Augustson-2016} between the hydrodynamic
and magnetic cases. It is worth mentioning that in the enhanced cases, the
amplitude of the anti-solar differential rotation is much weaker than
that from the original solar-like state. This is a direct consequence
of magnetic quenching of differential rotation \citep[see,
  e.g.][]{Brun2004,Brun-2005, Kapyla2017, Bice2023}. Our results are
in agreement with asteroseismic observations, as most of the observed
intermediate-mass MS stars ($M<9M_\odot$) have nearly solid body rotation
\citep{2014MNRAS.444..102K, 2021osvm.confE..27B, 2023Ap&SS.368..107B}.

Toroidal fields roughly comparable to the poloidal ones are found in
the statistically steady state at the stellar surface. These fields
are most likely transported from the enhanced core dynamo to the
surface due to turbulent diffusion. In the relaxed state the surface
magnetic field does not show signs of decay. In runs with Dipole B and
B* the surface field has typically between $94\%$ and $97\%$ of its
energy concentrated in the dipole mode. The relation between $\beta$
and $B_\mathrm{rms}$
is not prominent on the stellar surface (see right panels of
Fig.~\ref{fig:Brms-surf}), therefore, $\beta$ seems to affect mostly
the enhanced core dynamo. In Ap/Bp stars, the distribution of $\beta$
seems to be random, with no apparent relation to the surface dipolar
strength \citep{2007AN....328..475H, 2019MNRAS.483.3127S,
  Shultz2019}. The surface poloidal field strength is significantly
weaker in simulations with horizontal dipoles than in the rest of
cases.

The fossil fields were added after the saturation of the core dynamo.
Nevertheless, all of the current cases with a sufficiently strong
axial dipole land in a strong field branch where the dynamo mode
changes from predominantly multipolar to dipolar. This coincides with
results from other studies where the initial magnetic field was varied
similarly \citep[e.g.][]{Gastine-2012}.
In this work we explored very idealized tilted poloidal fields,
roughly based on the observed large-scale magnetic fields of Ap/Bp
stars. However,
fossil fields coming from the PMS evolution could in principle have
complex topologies. Unfortunately, very little is known about the
geometries and topologies of Herbig Ae/Be stars, the progenitors of
Ap/Bp stars \citep[see e.g.][]{2020AzAJ...15a..68H}. A logical next
step would be to model the PMS evolution of these stars, similarly to
\cite{2017ApJ...846....8E}, but concentrating on Herbig Ae/Be
stars. The obtained
magnetic fields will probably become mainly dipolar during the
main-sequence, as shown by \cite{Braithwaite-2006}, even random
initial magnetic fields can potentially relax into a dipolar
structure. However, how a more realistic configuration can affect an
existing core dynamo remains unknown.
 
\begin{acknowledgements}
JPH acknowledges financial support from ANID/DOCTORADO BECAS CHILE 72240057. DRGS gratefully acknowledges support by the ANID BASAL project FB21003 and the Alexander-von-Humboldt foundation. The simulations were performed with resources provided by
the Kultrun Astronomy Hybrid Cluster via the projects Conicyt Quimal
\#170001, Conicyt PIA ACT172033, and Fondecyt Iniciacion 11170268.
PJK acknowledges the stimulating discussions with participants of the
Nordita Scientific Program on ``Stellar Convection: Modelling, Theory
and Observations'' in August and September 2024 in Stockholm. CAOR acknowledges financial support from ANID (DOCTORADO DAAD-BECAS CHILE/62220030) as well as financial support from DAAD (DAAD/BECAS Chile, 2023 - 57636841 ). FHN acknowledges funding from the program Unidad de Excelencia Mar\'ia Maeztu, reference CEX2020-001058-M.
\end{acknowledgements}

%
\bibliographystyle{aa}
\bibliography{astro.bib}
%


\end{document}